\documentclass[final,3p,times,twocolumn]{elsarticle}
\usepackage[english]{babel}

\usepackage{soul}
\usepackage{graphicx}
\usepackage[colorlinks=true, allcolors=blue]{hyperref}
\usepackage{rotating}
\usepackage{multirow}
\usepackage{amssymb}
\usepackage{amsmath}
\usepackage{subfig}
\usepackage{lineno}
\usepackage{comment}
\usepackage{microtype}
\usepackage{hyperref}
\usepackage[table, dvipsnames]{xcolor}
\usepackage[export]{adjustbox}
\journal{}

\begin{document}

\begin{frontmatter}
\title{Status and future directions for direct cross-section measurements of the $^{13}$C($\alpha$,$n$)$^{16}$O reaction for astrophysics}
\author[add1]{L. Csedreki\corref{cor1}}
\ead{csedreki@atomki.hu}
\author[add1]{Gy. Gy\"urky}
\author[add4,add5]{D. Rapagnani}
\author[add7]{G.F. Ciani}
\author[add6]{M. Aliotta}
\author[add4,add5]{C. Ananna}
\author[add6]{L. Barbieri}
\author[add7,add15]{F. Barile}
\author[add8]{D. Bemmerer}
\author[add4,add5]{A. Best}
\author[add8]{A.~Boeltzig}
\author[add9]{C. Broggini}
\author[add6]{C.G. Bruno}
\author[add9,add10]{A. Caciolli}
\author[add11,add14]{F. Casaburo}
\author[add13]{F. Cavanna}
\author[add13,add12]{P. Colombetti} 
\author[add2,add3]{A. Compagnucci}
\author[add11,add14]{P. Corvisiero}
\author[add6]{T. Davinson}
\author[add16]{R. Depalo} 
\author[add4,add5]{A. Di Leva}
\author[add1,add26]{Z. Elekes} 
\author[add3]{F. Ferraro}
\author[add18]{A. Formicola}
\author[add1]{Zs. F\"ul\"op}
\author[add13,add12]{G. Gervino} 
\author[add16]{A. Guglielmetti} 
\author[add18]{C. Gustavino}
\author[add4,add5]{G. Imbriani}
\author[add3]{M. Junker} 
\author[add20,add25]{M. Lugaro} 
\author[add9,add10]{P. Marigo}
\author[add6]{J. Marsh}
\author[add8]{E. Masha} 
\author[add9]{R. Menegazzo} 
\author[add4,add5] {D. Mercogliano}
\author[add7]{V. Paticchio} 
\author[add7]{R. Perrino}
\author[add9,add10]{D. Piatti}
\author[add11,add14]{P. Prati}
\author[add6]{D. Robb}
\author[add7,add15]{L. Schiavulli}
\author[add6]{R. S. Sidhu}
\author[add9,add10]{J. Skowronski}
\author[add18,add23]{O. Straniero} 
\author[add1]{T. Sz\"ucs}
\author[add11]{S. Zavatarelli}

\cortext[cor1]{Corresponding author}

\address[add1]{HUN-REN Institute for Nuclear Research (HUN-REN ATOMKI), PO Box 51, 4001 Debrecen, Hungary}
\address[add2]{Gran Sasso Science Institute, Viale F. Crispi 7, 67100, L'Aquila, Italy }
\address[add3]{Istituto Nazionale di Fisica Nucleare Laboratori Nazionali del Gran Sasso (LNGS), Via G. Acitelli 22, 67100 Assergi, Italy}
\address[add4]{Universit\`a degli Studi di Napoli ``Federico II'', Dipartimento di Fisica ``E. Pancini'', Via Cintia 21, 80126 Napoli, Italy}
\address[add5]{Istituto Nazionale di Fisica Nucleare, Sezione di Napoli, Via Cintia 21, 80126 Napoli, Italy}
\address[add6]{SUPA, School of Physics and Astronomy, University of Edinburgh, Peter Guthrie Tait Road, EH9 3FD Edinburgh, United Kingdom}
\address[add7]{Istituto Nazionale di Fisica Nucleare, Sezione di Bari, Via E. Orabona 4, 70125 Bari, Italy}
\address[add8]{Helmholtz-Zentrum Dresden-Rossendorf, Bautzner Landstra\ss{}e 400, 01328 Dresden, Germany}
\address[add9]{Istituto Nazionale di Fisica Nucleare, Sezione di Padova, Via F. Marzolo 8, 35131 Padova, Italy}
\address[add10]{Universit\`a degli Studi di Padova, Via F. Marzolo 8, 35131 Padova, Italy}
\address[add13]{Istituto Nazionale di Fisica Nucleare, Sezione di Torino , Via P. Giuria 1, 10125 Torino, Italy}
\address[add12]{Universit\`a degli Studi di Torino, Via P. Giuria 1, 10125 Torino, Italy}
\address[add11]{Istituto Nazionale di Fisica Nucleare, Sezione di Genova, Via Dodecaneso 33, 16146 Genova, Italy}
\address[add14]{Universit\`a degli Studi di Genova, Via Dodecaneso 33, 16146 Genova, Italy}
\address[add15]{ Universit\`a degli Studi di Bari, Dipartimento Interateneo di Fisica, Via G. Amendola 173, 70126 Bari, Italy}
\address[add16]{Universit\`a degli Studi di Milano \& Istituto Nazionale di Fisica Nucleare, Sezione di Milano, Via G. Celoria 16, 20133 Milano, Italy} 
\address[add18]{Istituto Nazionale di Fisica Nucleare, Sezione di Roma, Piazzale A. Moro 2, 00185 Roma, Italy}
\address[add19]{Faculty of Physics, University of Warsaw, ul. Pasteura 5, 02-093 Warszawa, Poland}
\address[add20]{Konkoly Observatory, HUN-REN Research Centre for Astronomy and Earth Sciences, Konkoly Thege Miklós út 15-17, H-1121 Budapest, Hungary}
\address[add21]{Politecnico di Bari, Dipartimento Interateneo di Fisica, Via G. Amendola 173, 70126 Bari, Italy}
\address[add22]{ Technische Universit\"at Dresden, Institut f\"ur Kern- und Teilchenphysik, Zellescher Weg 19, 01069 Dresden, Germany}
\address[add23]{INAF Osservatorio Astronomico d'Abruzzo, Via Mentore Maggini, 64100 Teramo, Italy}
\address[add24]{Universit\`a degli Studi della Campania L. Vanvitelli, Dipartimento di Matematica e Fisica, Via Lincoln 5 - 81100 Caserta, Italy}
\address[add25]{ELTE Eötvös Loránd University, Institute of Physics and Astronomy, Pázmány Péter sétány 1/A, Budapest 1117, Hungary
}
\address[add26]{Institute of Physics, Faculty of Science and Technology, University of Debrecen, Egyetem tér 1., H-4032 Debrecen, Hungary}


\begin{abstract}
The $^{13}$C($\alpha$,$n$)$^{16}$O reaction 
is the main neutron source of the s-process taking place in thermally pulsing AGB stars and it is one of the main candidate sources of neutrons for the i-process in the astrophysical sites proposed so far. Therefore, its rate is crucial to understand the production of the nuclei heavier than iron in the Universe. For the first time, the LUNA collaboration was able to measure the $^{13}$C($\alpha$,$n$)$^{16}$O cross section at $E_{\rm{c.m.}}$=0.23$-$0.3 MeV drastically reducing the uncertainty of the $S(E)$-factor in the astrophysically relevant energy range. 
In this paper, we provide details and critical thoughts about the LUNA measurement and compare them with the current understanding of the $^{13}$C($\alpha$,$n$)$^{16}$O reaction in view of future prospect for higher energy measurements. The two very recent results (from the University of Notre Dame and the JUNA collaboration) published after the LUNA data represent an important step forward. There is, however, still room for a lot of improvement in the experimental study of the $^{13}$C($\alpha$,$n$)$^{16}$O reaction, as emphasized in the present manuscript. We conclude that to provide significantly better constraints on the low-energy extrapolation, experimental data need to be provided over a wide energy range, which overlaps with the energy range of current measurements.
Furthermore, future experiments need to focus on the proper target characterisation, the determination of neutron detection efficiency having more nuclear physics input, such as angular distribution of the $^{13}$C($\alpha$,$n$)$^{16}$O reaction below $E_{\alpha}<$0.8 MeV and study of nuclear properties of monoenergetic neutron sources and/or via the study of sharp resonances of $^{13}$C($\alpha$,$n$)$^{16}$O. Moreover, comprehensive, multichannel R-matrix analysis with a proper estimate of uncertainty budget of experimental data are still required. 

\end{abstract}
\end{frontmatter}

\section{Introduction} \label{sec:introduction}

Half of the chemical elements heavier than iron in the Universe are produced in stars via slow neutron captures (the s-process) through sequences of neutron capture reactions and $\beta$ decays. Spectroscopic observations combined with stellar models support the $^{13}$C($\alpha$,$n$)$^{16}$O reaction as the neutron source for s-process in low-mass Thermally Pulsing Asymptotic Giant Branch (TP-AGB) stars \cite{Staniero2006, Lugaro2023}. 
A TP-AGB star is composed of a degenerate carbon-oxygen core surrounded by a thin He-rich shell and an extended, convective H-rich envelope. Periodically, these stars undergo thermonuclear instabilities caused by flashes of He-burning shell, called thermal pulses (TPs). Each He-flash generates a convective zone that mixes the carbon produced by the triple-$\alpha$ reaction up to the top of the He shell. During the TP, the shell H burning is extinguished, while after a TP the external convection may penetrate the He-rich shell. Because of this recurrent mixing process, so-called third dredge up, carbon enriched material is brought up to the stellar surface. As a byproduct of these recursive mixing episodes, a so-called $^{13}$C-pocket can be formed at the top of the He shell through the reaction sequence $^{12}$C(p,$\gamma$)$^{13}$N ($\beta$$^{+}$,$\nu$)$^{13}$C \cite{Staniero2006, Gallino1998}.
In between the TP events (referred later as an interpulse period), the $^{13}$C($\alpha$,$n$)$^{16}$O reaction is activated at temperatures of about 90 MK and provides a neutron flux with relatively low density of around 10$^{7}$ neutrons/cm$^{3}$ for about 10$^{4}$ years. Starting from Fe-group nuclei, neutron capture reactions followed by $\beta$-decays along the valley of stability of nuclides \cite{Kaeppeler:2011} produce heavier chemical elements in the He-rich region reaching up to $^{209}$Bi, the highest mass stable nucleus. To predict chemical abundances in TP-AGB, the complex study of stellar structure, composition and mixing phenomena needs to be supported by accurate and precise reaction rates derived from reaction cross sections in the relevant astrophysical energy window.

Furthermore, the $^{13}$C($\alpha$,$n$)$^{16}$O cross section around $E_{\rm{c.m.}}$=0.14$-$0.25 MeV is crucial to estimate the energy balance and neutron flux in the thermal pulses for those stellar models where the $^{13}$C nuclei can survive the interpulse period, be ingested into the convective shell, and burn at higher temperature of around 200 MK \cite{Cristallo_2009, Bisterzo2015}. 
It is also significant for one of the astrophysical scenarios assumed to explain the observed surface abundances of a fraction of the Carbon Enhanced Metal Poor (CEMP) stars, post-AGB stars, and of Sakurai's object \cite{Hampel_2016, Hampel_2019, Herwig_2011}, where protons are ingested directly into a thermal pulse. This mixing results in the production of $^{13}$C nuclei that ignite the $^{13}$C($\alpha$,$n$)$^{16}$O reaction at temperatures around 200 MK, providing a neutron flux with a neutron density up to 10$^{14}$ neutrons/cm$^{3}$, and driving the intermediate neutron capture, so-called, i-process, i.e., neutron capture further away from the valley of beta stability \cite{Lugaro2023}. The associated Gamow energies in this case are $E_{\rm{c.m.}}$=0.2$-$0.54 MeV. 

The $^{13}$C($\alpha$,$n$)$^{16}$O reaction can become an important source of background in ultra-low event rate research, such as underground nuclear, neutrino research and double-beta decay experiments \cite{Arnquist2022, COOLEY2018, Febbraro2020}. In fact the $\alpha$-emitters of the $^{232}$Th and $^{238}$U decay chains combined with an environment of carbon-rich construction material, e.g., plastic for neutron shielding or electrical insulation, and the detector materials of organic scintillators can become a significant contributor to neutron background, mimicking neutrino signal.

To constrain the phenomena described  above, an accurate knowledge of the total and differential reaction cross section ($\sigma(E)$ and $d\sigma(E)/d\Omega(E)$, respectively), over a wide energy range is crucial. These quantities can be extracted, e.g., from the experimental yield ($\textit{Y}$), defined as the number of neutrons per charge, and expressed by the formula:

\begin{equation}
\label{eq:yield}
Y=\eta(E_{n})\int^{E_{\rm \alpha}}_{E_{\alpha}-\Delta E}\frac{\sigma(E)}{\epsilon_{\rm{eff}}(E)}dE \quad ,
\end{equation}
 
where $\eta(E_{n})$ is the neutron detection efficiency at a given neutron energy, $E_{\alpha}$ is the beam energy and $\epsilon_{\rm{eff}}(E)$ is the effective stopping power. $\Delta$$E$ is the energy lost by the beam in the target.
Since the measured yield is proportional to the integral of $\sigma(E)$/$\epsilon_{\rm{eff}}(E)$, the cross section needs to be extracted by an iterative approach and the $\sigma(E)$ can be converted to so-called astrophysical $S(E)$-factor defined by:

\begin{equation}
\label{eq:s_faktor}
\sigma(E)=\frac{1}{E}\exp^{-2\pi\eta}S(E),
\end{equation}
where $\eta$ represents the Sommerfeld parameter \cite{Iliadis2007}. Each parameter to obtain the cross section requires careful and specific evaluation of the corresponding uncertainties.

Here, we summarize the experimental parameters of the LUNA measurement \cite{Ciani2021, Ciani2020, Csedreki2021, BALIBREA2018} and discuss in details the efforts made to keep under control the systematic effects due to target characterization. This is one of the main, and often overlooked, source of uncertainty for $^{13}$C($\alpha$,$n$)$^{16}$O cross-section measurements.
We also guide the way forward to an improved determination of the cross section. 
Section \ref{sec:13an_reaction} describes the state of the art of cross-section measurements of the $^{13}$C($\alpha$,$n$)$^{16}$O reaction; Section \ref{sec:Exp_parameters} provides the details of the experimental parameters affecting the uncertainty budget of cross-section data. Section \ref{sec:overview} gives details for the selection of experimental data sets to the LUNA extrapolation of the cross section to Gamow energies. Section \ref{sec:R_matrix} provides the work of LUNA for low-energy extrapolation and the comparison with more recent measurements of the $^{13}$C($\alpha$,$n$)$^{16}$O reaction.
The conclusions and prospects of upcoming experiments are discussed in Section \ref{sec:summary}.

\section{The $^{13}$C($\alpha$,$n$)$^{16}$O cross section: state of the art} \label{sec:13an_reaction}

\begin{figure}[htb!]
    \centering
    \includegraphics[width=\columnwidth]{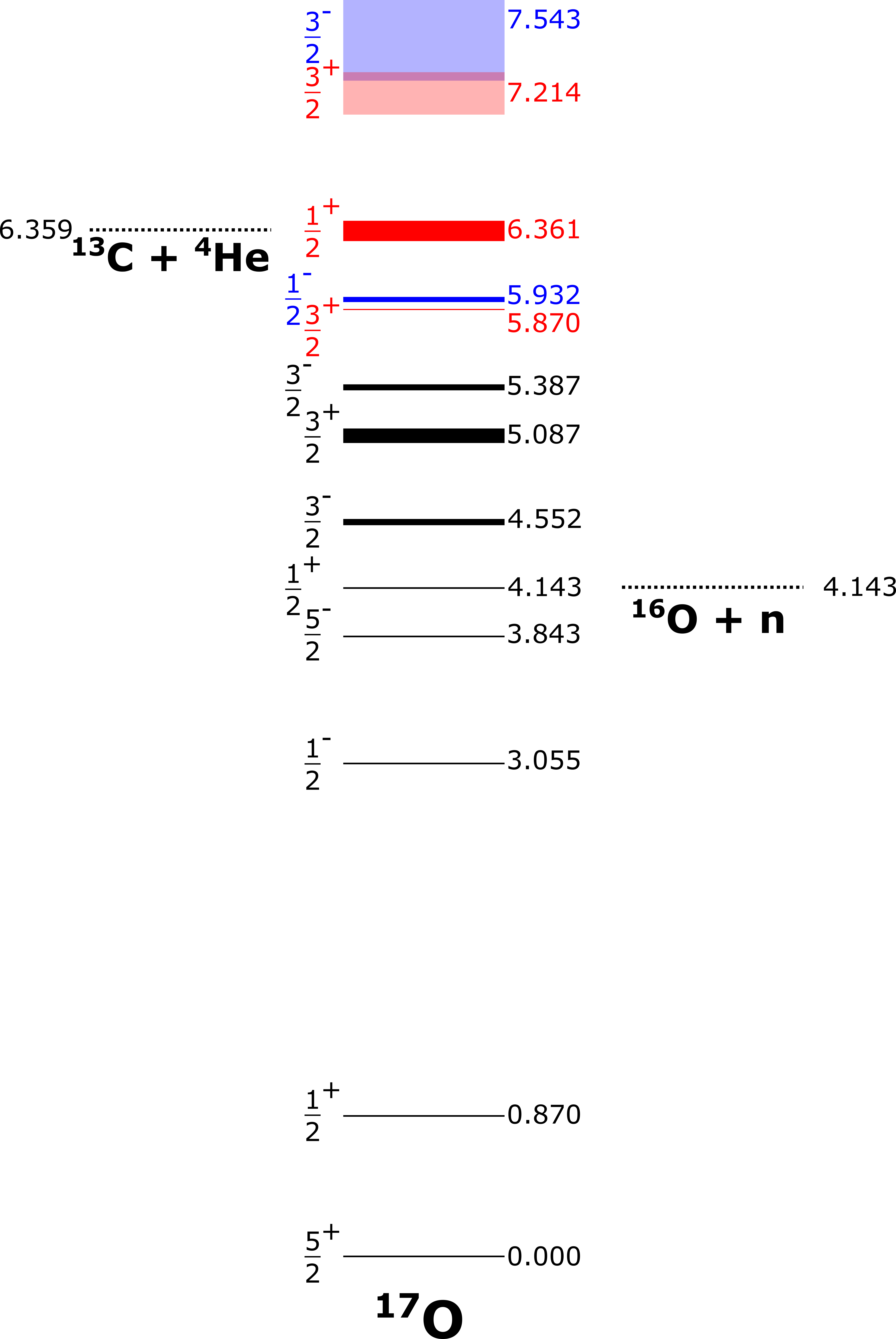}
    \caption{(Colour online) Level scheme of $^{17}$O. Widths of the horizontal lines reflect the widths of the excited states. 
    The sharp 5/2 levels with altering parities at 6.861, 7.166, 7.379 and 7.382 MeV are omitted, because their effect over the cross section at stellar energies is marginal.}
    \label{fig:level_scheme}
\end{figure}

The low-energy behaviour of the $^{13}$C($\alpha$,$n$)$^{16}$O reaction cross section is determined by several resonances and their interference patterns \cite{Ciani2021, Heil2008, Faestermann2015, Tilley1993}. Thus the low-energy extrapolation is affected by the uncertainties of the resonance parameters at $E_{\alpha}$ = $-$ 641 keV ($J^{\pi}$ = 3/2$^{+}$), $E_{\alpha}$ = $-$ 569.7 keV ($J^{\pi}$ = 1/2$^{-}$), $E_{\alpha}$ = $-$ 2.7 keV ($J^{\pi}$ = 1/2$^{+}$) and $E_{\alpha}$ = 1162.6 keV ($J^{\pi}$ = 3/2$^{+}$).
In particular, the tail of the broad near-threshold resonance with J$^{\pi}$ = 1/2$^{+}$ and total resonance width (equal with the neutron partial width) $\Gamma_{total}$ = $\Gamma_{n}$ = (136 $\pm$ 5) keV is found to cause an increase of the $S(E)$-factor below $E_{\rm{c.m.}}<$0.3 MeV \cite{Faestermann2015}. Table \ref{table:resonance_parameter} summarizes the parameters of the resonances in $^{13}$C($\alpha$,$n$)$^{16}$O relevant for the $S(E)$-factor extrapolation below $E_{\rm{c.m.}}<$0.8 MeV and Figure \ref{fig:level_scheme} shows the $^{17}$O level scheme.

\begin{table}[t!]
\caption{Parameters of the resonances (resonance energy $E_{r}$, excitation energy $E_{x}$ and Lorentzian width $\Gamma_{n}$) relevant to the $S(E)$-factor extrapolation of $^{13}$C($\alpha$,$n$)$^{16}$O at $E_{\rm{c.m.}}<$0.8 MeV. All values are given in keV and the source references are provided.}
\begin{center}
\resizebox{1\columnwidth}{!}{
\begin{tabular}{|c | c | c | c | c | c | c | c |}
\hline
$J^{\pi}$	&	$E_{r}$ & \multicolumn{2}{c}{\cite{Tilley1993}} & \multicolumn{2}{c}{\cite{Heil2008}} & \multicolumn{2}{c}{\cite{Faestermann2015}} \\ 
 & & $E_{x}$ & $\Gamma_{n}$  & $E_{x}$ & $\Gamma_{n}$  & $E_{x}$ & $\Gamma_{n}$ \\ \hline
3/2+ & -641 & 5869.62 & 6.6 & 5868.4 & 25.2 & 5869.7  &  \\
1/2- & -569.7 & 5931.6 & 32 & 5923.2 & -48.1 & 5931 & 33   \\
1/2+ & -2.7 & 6361.5 & 124 & 6379.5 & 158.1 & 6363.4 & 136 \\
5/2- & 1053.9 & 7165.86 & 1.38 & 7164.6  & 1.88 & 7165.4 &  \\
3/2+ & 1162.6 & 7214 & 263 & 7247.7 & 340.1 & 7216 & 262 \\
5/2+ & 1332.9 & 7379.23 & 0.61 & 7377.9 & 0.41 & 7380.1  & \\
5/2- & 1336.5 & 7382.37 & 0.9 & 7380.7 & 1.77 & 7380.1 & \\
3/2- & 1548.2 & 7543 &  500 & & & & \\
7/2+ & 1590 & 7573.5 & $<$ 0.1 & & & 7573.5 &  \\
 \hline
\end{tabular}
\label{table:resonance_parameter}
}
\end{center}

\end{table}

Despite their marginal effect in the $S(E)$-factor extrapolation, sharp resonances between $E_{\rm{c.m.}}$ = 0.7$-$1.2 MeV can be used to constrain experimental parameters, such as target thickness and neutron detection efficiency \cite{PEREIRA2010}. Therefore, their parameters (except the resonance strength $\omega\gamma$) are also indicated in Table \ref{table:resonance_parameter}. It is worth mentioning that parameters of sharp resonances show discordance in literature, e.g., the works of \cite{Bair_Haas1973, Brune1993} and \cite{Ramström1976, Ru2023} suggest $\approx$12 and 17 eV, respectively, for the $\omega\gamma$ of the $E_{\rm{c.m.}}\approx$0.8 MeV resonance. Although Ref. \cite{Ru2023} attributes this discrepancy to the contribution of the non-resonant cross section at the resonance energy and the stopping power of $\alpha$ particles in carbon, a dedicated measurement of the resonance parameters is advisable to confirm that.

Because of the 2.215 MeV \cite{Wang2021} Q-value of the $^{13}$C($\alpha$,$n$)$^{16}$O reaction, at beam energies $E_{\rm{c.m.}}$=0.23$-$1.53 MeV, neutrons are emitted over a wide energy range $E_{n}$= 2.1$-$4.1 MeV, depending also on the emission angle $\theta$. The neutron angular distribution will be discussed in Section \ref{subsec:neutron_eff}.

To properly estimate the $\alpha$-induced neutron background generated on $^{13}$C nuclei, the reaction cross section needs to be determined up to several MeV range. 

For example, $^{210}$Po with the emission of $E_{\rm{c.m.}}$=4.0 MeV initiates the $^{13}$C($\alpha$,$n$)$^{16}$O reaction, in turn, opening the $\alpha$, $n_{0,1,2}$ branches and populating the $E_{x}$=0 ($n_{0}$), 6.05($n_{1}$) and 6.13($n_{2}$) MeV states in $^{16}$O, respectively \cite{Peters2017, Febbraro2020}. Therefore, the $^{13}$C($\alpha$,$n$)$^{16}$O reaction has been extensively studied by past and recent experiments over a wide energy range between $E_{\rm{c.m.}}$= 0.23$-$6.1 MeV. 

\begin{figure*}[htb!]
    \centering
    \includegraphics[width=2\columnwidth]{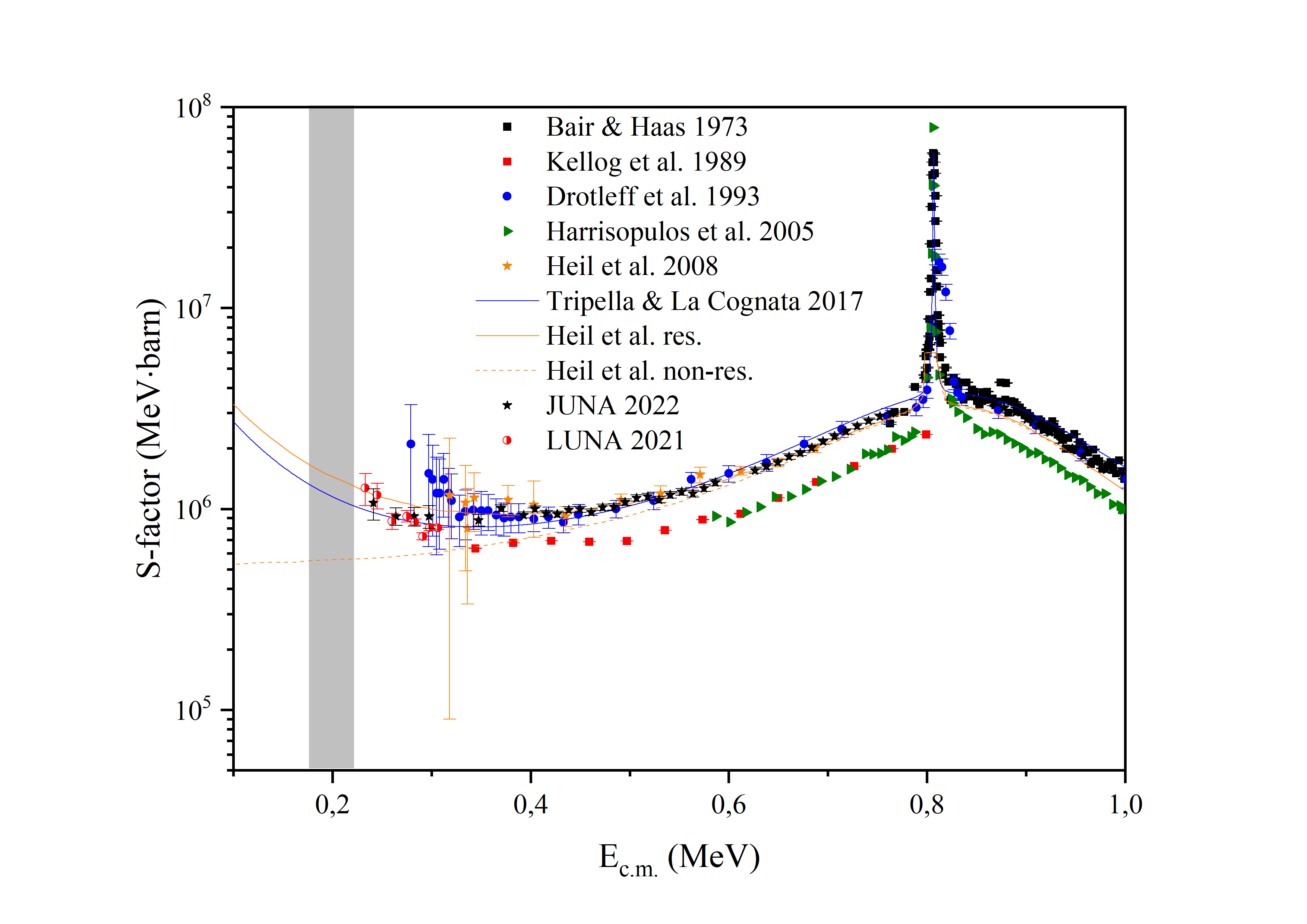}
    \caption{(Colour online) Experimental $S(E)$-factor data of the $^{13}$C($\alpha$,$n$)$^{16}$O reaction combined with theoretical calculations with and without the effect of the near-threshold resonance. The grey area represents the Gamow-window of the reaction for about T=0.1 GK.}
    \label{fig:Exp_S_factor_data}
\end{figure*}

Experimental $S(E)$-factor data of $^{13}$C($\alpha$,$n$)$^{16}$O \cite{Bair_Haas1973,Kellog1989,Drotleff1993, Harissopulos2005, Heil2008, Ciani2021, Gao2022, deBoer2024}, combined with theoretical calculations \cite{Trippella&LaCognata2017, Heil2008} with and without the effect of the near-threshold resonance, are shown in Figure~\ref{fig:Exp_S_factor_data}. 

Beside direct measurements, indirect studies were performed with the goal of determining the energy of the near threshold levels by $\gamma$-spectroscopy \cite{Faestermann2015}; the squared Coulomb-modified asymptotic normalization coefficient (ANC) \cite{Johnson2006, Avila2015}; or the similar spectroscopic factor \cite{Kubuno2003, Keeley2003, Pellegriti2008, Guo2012, Mezhevych2017}. The latter were used to calculate the low-energy astrophysical $S(E)$-factor and the $^{13}$C($\alpha$,$n$)$^{16}$O reaction rate.
Ref. \cite{deBoer2020} reported that the $S(E)$-factor calculation based on the Coulomb renormalized $\alpha$-particle ANC ($\overset{\sim}{\rm C}$) obtained in different indirect measurements bears roughly 20\% uncertainty, which is also highlighted by Ref. \cite{hebborn2023}, and the inconsistency of direct measurements gives another $\sim$40\% to the uncertainty budget in the s-process temperature range. The significant effect of such uncertainties in different astrophysical scenarios was further emphasized by Ref. \cite{Cristallo2018}.

In the last three years new experimental data has been published, which potentially reduce the uncertainties of experimental cross-section data. Direct measurements performed underground by the LUNA collaboration \cite{Ciani2021} in the $E_{\rm{c.m.}}$=0.23-0.31 MeV region using the thin target approach and the work done by the JUNA collaboration \cite{Gao2022} in $E_{\rm{c.m.}}$=0.24-1.90 MeV using thick and thin targets, have allowed to significantly reduce the uncertainty below $E_{\alpha}<$ 0.40 MeV providing improved constraint to calculate s-process nucleosynthesis. 
Together with a recent measurement above ground with high angular resolution in the $E_{\rm{c.m.}}$=0.61-4.8 MeV Ref. \cite{deBoer2024} reports that based on their R-matrix analysis using 714 angular distributions and the combination of Bayesian uncertainty estimation, the uncertainty of the extrapolation of cross section significantly reduced to the level of $\sim$5\% over the entire Gamow energy range.

\section{Experimental aspects} \label{sec:Exp_parameters}

In this section, we review the experimental parameters affecting the determination of the cross section of $^{13}$C($\alpha$,$n$)$^{16}$O towards its astrophysically relevant energy region with a focus on the LUNA experiment. 
The most important experimental parameters of the direct cross-section measurements are summarized in Table ~\ref{tab:exp_data_summary}.

\subsection{Experimental apparatus and its background} \label{subsec:env_background} 

The LUNA collaboration performed the study of $^{13}$C($\alpha$,$n$)$^{16}$O at the deep-underground accelerator LUNA400 \cite{FORMICOLA2003,pantaleo2021,piatti2022} at the Gran Sasso National Laboratory (LNGS). A detailed discussion on the origin of background contributions and the advantage of deep-underground location for rare event research is found in the literature \cite{LUNA2018, BEST2016, Iliadis2007, 2020PhRvD.101l3027G}. Here we just emphasize that due to the rock overburden of LNGS, the background neutron flux is reduced by up to 4 orders of magnitude with respect to the neutron flux measured on the surface of the earth.

In general, experiments have used multiple approaches to detect neutrons. Gas filled proportional counters based on $^{3}$He gas \cite{Csedreki2021, Drotleff1993,Brune1993, Harissopulos2005}; $^{10}$BF$_{3}$ gas \cite{Bair_Haas1973, Ramström1976} or their combination \cite{Brandenburg2023} are widely used. Although these provide higher neutron detection efficiency, direct information on neutron energies is lost by the moderation of neutrons. Thus, the discrimination of background events is more challenging. 
Experimental setups based on scintillators, e.g., \cite{Davids1968, Febbraro2020, deBoer2024} allow angular distribution measurement to be performed, but their neutron detection efficiency is limited.
The experimental study of the angular distribution of the emitted neutrons is limited by the thermalization process in the moderator material using gas filled proportional counters. Thus, simulation of the angular distribution must be considered during the extraction of the cross section. This issue is further discussed in Section \ref{subsec:neutron_eff}.

The location of the experimental apparatus of LUNA and the selected stainless steel material of the enclosure of the $^{3}$He counters imply a unique low environmental background with 3.3 counts/hour counting rate in the detector. Although a very low level of environmental background is achieved, the reaction yield of the $^{13}$C($\alpha$,$n$)$^{16}$O reaction drops to the 1 event/hour level at energies close to the s-process Gamow peak. Therefore, background is still one of the most severe sensitivity limitations of the experimental apparatus. \\

To further improve the signal-to-noise ratio, the LUNA collaboration used Pulse Shape Discrimination (PSD) technique based on digital filter to convert the integrated signal of the charge sensitive preamplifiers to a current pulse. The application of this PSD technique allows suppression of the internal $\alpha$-induced background by\> 98.5\% and thus reduces the total background of the LUNA neutron array to (1.23$\pm$ 0.12) counts/hour \cite{BALIBREA2018,Badala2022}. 
Note that PSD also reduces the neutron detection efficiency, and should therefore be considered carefully in the cross-section determination.
The achieved background rate by LUNA represents an improvement of two orders of magnitude over similar setups \cite{Drotleff1993, Harissopulos2005} used in the past and it is a factor of 4 better than the value obtained recently by the JUNA collaboration (see Table \ref{tab:exp_data_summary}).

The uncertainty due to the background of the setup can be eliminated using indirect detection of neutron as described by Heil et al. \cite{Heil2008}. These authors used a 4$\pi$ BaF$_{3}$ calorimeter with n/$\gamma$ converter based on spherical cadmium-loaded paraffin to achieve a practically background-free condition. With this technique, low-energy gammas are well separated from the signal of the $\gamma$ cascade from the $^{113}$Cd(n,$\gamma$)$^{114}$Cd reaction yield at a total energy of 9043 keV, the drawback is that the total neutron detection efficiency is limited by the n/$\gamma$ conversion factor. 

\subsection{Neutron detection efficiency}  \label{subsec:neutron_eff}

In contrast to $\gamma$-ray spectroscopy, the determination of the neutron detection efficiency as a function of neutron energy $\eta$(E$_{n}$) is challenging mainly due to the limited choices of sources with accurately known energy spectra and/or angular distributions (in the case of nuclear reactions based source) and, in some cases, to the limited availability of accurately calibrated sources. A standard procedure is to employ radioactive sources ($^{252}$Cf, AmBe), which emit neutrons with a continuous energy spectrum, in combination with Monte Carlo simulations \cite{Drotleff1993, Harissopulos2005, Denker1995}. This approach can be made more robust using nuclear reactions, e.g., the $^{51}$V(p,n)$^{51}$Cr reaction \cite{Deconninck1969, Falahat2013, Wrean2000, PEREIRA2010, Csedreki2021, Yu-Tian2022}. To constrain the uncertainty of the efficiency determination, the design of neutron detection setup should be optimized to obtain a relation between the detection efficiency and neutron energy as flat as possible along the energy range of interest \cite{PEREIRA2010}. A list of the widely used neutron emitters (both radioactive sources and nuclear reactions) is presented in Table ~\ref{tab:neutron_emitters}.

\begin{table}[htb!]
\caption{Properties of some neutron emitting radioactive sources and nuclear reactions often used for neutron detection efficiency determination.}
\begin{center}
\resizebox{1\columnwidth}{!}{
\begin{tabular}{|c | c | c | c | }
\hline
Neutron source &	Q value (MeV) &	Mean energy (MeV) &	Half life \\ \hline
	
$^{nat}$UO	& & & \\		
$^{239}$PuBe	& - & 4.6  &		24110(30) y\\
$^{241}$AmBe		& - & 4.0  &	432.6(6) y\\
$^{241}$AmLi	& - & 0.8  & 432.6(6) y \\	
$^{252}$Cf	& - & 2.13*  &		2.645(8) y \\
$^{88}$Y-Be	& & 0.152, 0.949 (0.5\%) ** &		106.626(21) d\\
$^{124}$Sb-Be & & 0.023  &			60.20(3) d\\ \hline
$^{2}$H($\gamma$,n)p	& -2.23 	& & \\	
D(d,n)$^{3}$He &	3.27 		& & \\
$^{7}$Li(p,n)$^{7}$Be &	-1.64 		& & \\
$^{9}$Be($\gamma$,n)$^{8}$Be	& -1.66 & & \\
$^{9}$Be(d,n)$^{10}$B	& 4.362  & & \\
$^{51}$V(p,n)$^{51}$Cr &	-1.535 	& &  \\
$^{57}$Fe(p,n)$^{57}$Co*** & -1.6186  & & \\
\hline
\end{tabular}
\label{tab:neutron_emitters}
}
\end{center}
\footnotesize $*$ From Ref.\cite{Fröhner1990}. Instead,  Ref. \cite{Brune1993} and \cite{Harissopulos2005} indicate 2-3 MeV as an average neutron energy\\
\footnotesize $**$ From Ref.\cite{Knoll2010}\\
\footnotesize $***$ From Ref.\cite{HUNT1978}\\

\end{table}		

\begin{figure}[hbt!]
\begin{center}
\begin{tabular}{c c}
\includegraphics[width=0.96\columnwidth]{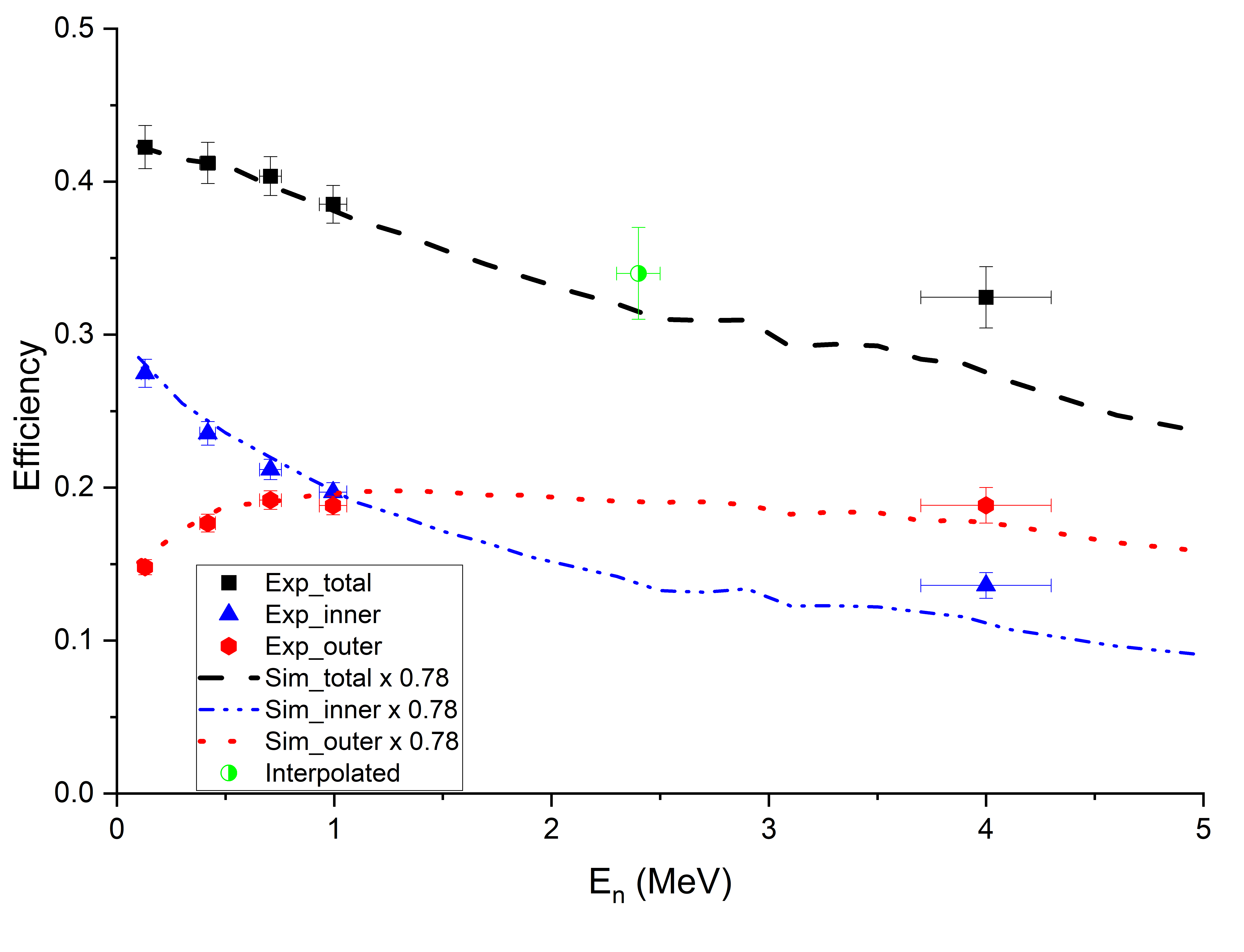}\\
\includegraphics[width=0.96\columnwidth]{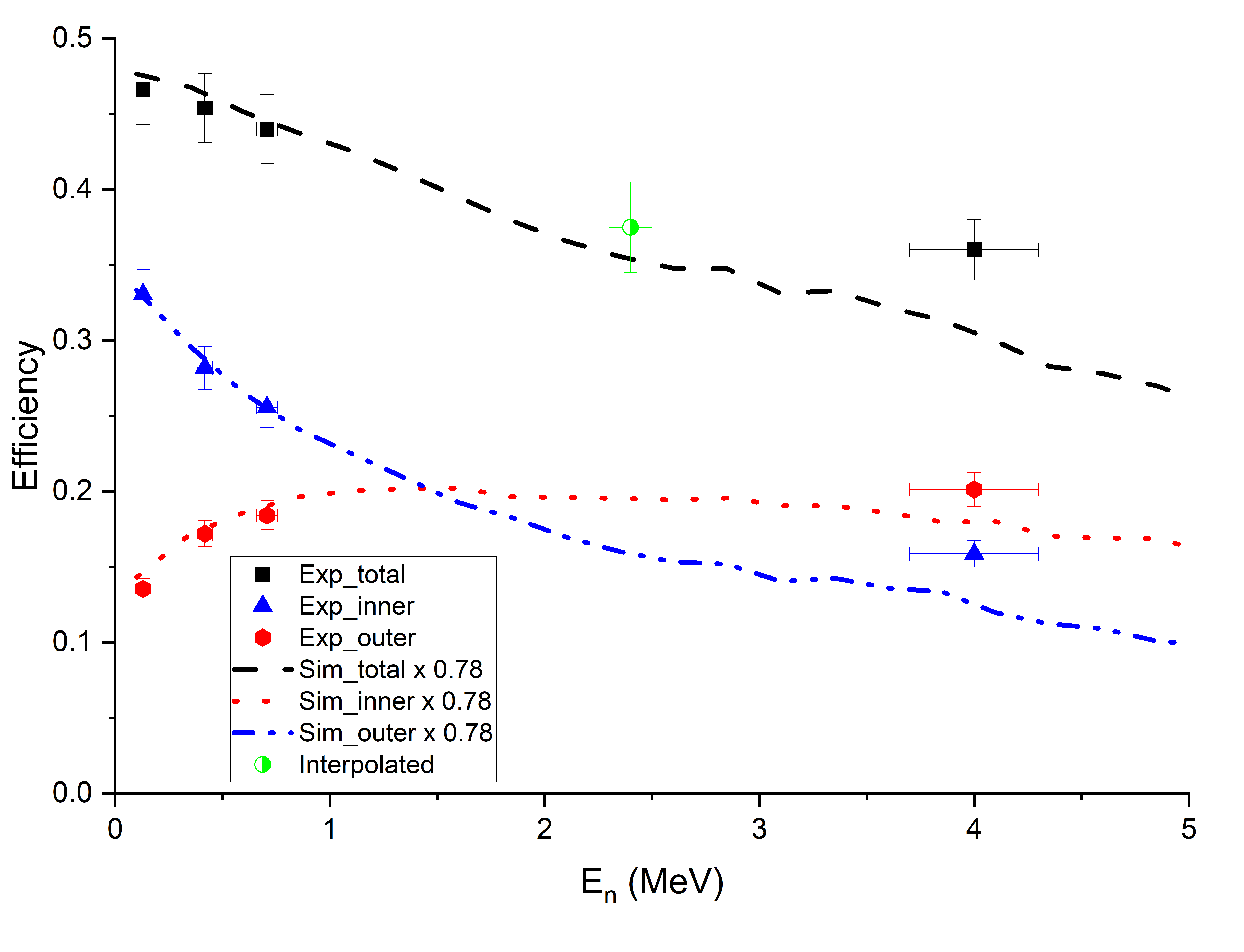}\\
\end{tabular}
\end{center}
\caption{(Colour online) Experimental efficiencies (filled symbols) and the rescaled simulated efficiency curve (dashed line) obtained using the vertical (upper panel) and horizontal (lower panel) setups. The simulated and the experimental efficiencies related to the inner (green squares and dotted lines) and outer (blue triangles and dash-dotted lines) rings of the setups are also presented. The interpolated efficiency value at $E_{\text{n}}$=2.4\,MeV are shown as green half empty dot.}
\label{fig:comparison}
\end{figure}

In the LUNA experiment, the $\eta$(E$_{n}$) was experimentally determined over neutron energy range $0.1-4.0$ MeV using the $^{51}$V(p,n)$^{51}$Cr reaction and an AmBe neutron source, combined with a detailed simulation implemented in the GEANT4 code.
The GEANT4 \footnote{Specifically the GEANT4 version 10.03, including the neutron high precision physics and thermal scattering corrections enabled for water and polyethylene.} toolkit \cite{Allison2016,Agostinelli2003} was used to simulate the detector response of the LUNA neutron array. 
This simulation was compared with the experimental data. The efficiency curve obtained from GEANT4 is shown in Figure~\ref{fig:comparison} together with the experimental results. The plotted experimental data were corrected for the kinematic energy distribution effect and angular distributions to obtain the nominal efficiency values. 
To cross-check the consistencies of the simulation, the results for the inner and outer rings of the detector array are shown separately.

Absolute neutron detection efficiencies of (37$\pm$3)\% (horizontal) and (34$\pm$3)\% (vertical) \footnote{Equal with a 8\% relative uncertainty. Uncertainty budget is described in Ref. \cite{Csedreki2021}} of the two setups were obtained in the $E_{\text{n}}$=2.2-2.6 MeV range, which corresponds to the energies of the emitted neutrons in the LUNA experiment. 
The efficiency was calculated as an average of the GEANT4 simulation and the linear interpolation of the experimental data as shown in Figure~\ref{fig:comparison}.

A discrepancy was found between the experimental and the simulated data. Therefore, a scaling factor of $0.78\pm0.01$ was applied to the simulated efficiency curves shown in Figure~\ref{fig:comparison} in order to match the experimental data. The presence of such a discordance between experimental and simulated efficiencies is rather general in literature \cite{Csedreki2021, PEREIRA2010, Yu-Tian2022, Falahat2013}. 
This may indicate a model dependent uncertainty of the method, which can be related to the uncertainty of angular distribution of nuclear reactions, geometrical effect of the setup, and/or physics input of MC simulation (e.g., scattering cross sections, molecular vibrational and rotational excitation modes in the moderator materials, etc.) \cite{Kittelmann2014, KNEIssl2019, Li2017, MARINO2007611, Li2018, Ende2016, Lemrani2006}. These sources of uncertainties are discussed in the following paragraphs.

As mentioned in Section \ref{sec:13an_reaction}, the $^{13}$C($\alpha$,$n$)$^{16}$O cross section may have significant energy-dependent anisotropy. Angular distribution measurements can be found in literature over a wide $E_{\alpha}$ region \cite{Walton1957, Schiffer1957, Febbraro2020, Prusachenko2022, deBoer2024}. 
Walton et al.~\cite{Walton1957} used a scintillator detector, which was placed at a distance from the thin target ($\Delta$$E<$20 keV) of 11.5 cm above $E_{\alpha}$= 2 MeV and 6.4 cm below this energy. Data was taken between 0° and 150° at angular intervals of 10° and at 85°, 95°, and 155°, acquiring enough statistics to keep the uncertainty below 5\%. A long counter was also used to obtain absolute cross-section data at $\theta$=0°, 29° and 146° at $E_{\alpha}$=0.8$-$3.5 MeV.
Moreover, Prusachenko et al.~\cite{Prusachenko2022} recently published angular distribution data in the E$_{\alpha}$=2-6.1 MeV region using a p-terphenyl scintillator and time-of-flight (ToF) technique.
Febbraro et al. \cite{Febbraro2020} performed measurement of $^{13}$C($\alpha$,$n$)$^{16}$O cross section in $E_{\alpha}$=4.2$-$6.4 MeV using two deuterated scintillator detectors for neutron detection and a single HPGe for $\gamma$-ray detection of $^{13}$C($\alpha$,$n$$\gamma_{2}$)$^{16}$O. Similarly, deBoer et al. \cite{deBoer2020} used deuterated scintillator array (ODeSA) \cite{Febbraro2019} to provide high angular resolution differential cross-section data between $\theta$=0° and 157.5° in $E_{\alpha}$=0.8$-$6.5 MeV. They could not measure the excited state cross section due to the background of fluorine and the resolution of their unfolding technique.

The angular distribution can also be calculated assuming pure radiation in a two-step process as described, e.g., by Ref. \cite{Iliadis2007} (Appendix D.2.) near sharp resonances. Moreover, R-matrix calculation can be used to obtain the angular distribution for a given $\alpha$-energy \cite{deBoer2020} based on experimental elastic scattering and nuclear reaction data. This effect is discussed in literature purely in terms of the uncertainty of cross-section data. Based on theoretical calculation, Ref. \cite{Gao2022} estimates a relative  deviation between the detection efficiency lower than 2.3\% assuming isotropic and anisotropic angular distribution in $E_{\alpha}$=0.3$-$0.8 MeV. This uncertainty can increase significantly near sharp resonances and at higher energies.\\ 
In the case of the LUNA measurement, due to the larger solid angle and low energies, the uncertainty implied by anisotropic angular distribution is far below the quoted relative 8\% assigned to $\eta$(E$_{n}$). 
Ref. \cite{Brandenburg2023} reports $\approx$ 10\% and $\approx$ 14\% at $\alpha$ energies below and beyond 5 MeV, respectively, using R-matrix method and Hauser-Feshbach calculation \cite{Mohr2018}. 

As a conclusion, the angular distribution of the $^{13}$C($\alpha$,$n$)$^{16}$O reaction is rather well known and supported by experimental data above $E_{\alpha}$ $=$ 0.8 MeV, however the absence of experimental data in the lower energy regime still requires theoretical predictions. Dedicated measurements, e.g., with a long counter setup and/or plastic scintillators using pulsed beam and ToF technique could provide data with the required precision. \\

In the context of low-energy nuclear astrophysics measurements, the MCNP, GEANT4 and FLUKA simulation codes are widely used to obtain parameters for cross-section calculations.
To study the precision of different simulation codes, van der Ende et al. \cite{Ende2016} performed a systematic study of boron-lined neutron detector characterisation using MCNPX and GEANT4. They confirmed the reliability of GEANT4 to accurately characterise a thermal neutron detector through the use of a special thermal elastic scattering matrix S($\alpha$,$\beta$) tables for neutron energies lower than 4 eV, as well as through the use of neutron high precision models.\footnote{From the QGSP$\_$BERT$\_$HPphysics package (G4NeutronHPElastic, G4NeutronHPInelastic, and G4NeutronHPCapture)} 

Related to the uncertainties implied by the geometry of the setup, Ref. \cite{Yu-Tian2022} suggests a solution to resolve the discrepancy by adding boron to the moderator material to mimic the neutron absorption effect.
Nevertheless, the uncertainty of geometry propagates negligible effect (relative $<$1\%) in $\eta(E_{\text{n}})$ in the LUNA experiment.

Finally, uncertainties of cross sections (e.g., neutron capture, thermal scattering) as inputs of the simulation and their effect over the simulated detection efficiencies are poorly discussed in literature. Especially, their energy dependent effect would require special attention. It is worth noting that this source of uncertainty also becomes significant in the case of measurement using scintillator detectors, where neutron scattering on the construction materials implies to apply even a $\sim$30\% correction \cite{deBoer2024}.

In this section, we discussed the different aspects of the determination of the neutron detection efficiency. 
In conclusion, dedicated studies of the scaling factor in a simple neutron detection geometry combining the use of more extended monoenergetic neutron sources in wide energy range are still advisable. A possible way to access this is the study of branching of $^{51}$V(p,n)$^{51}$Cr above E$_{n}$$>$1 MeV through the detection of gammas from excited states in $^{51}$Cr (e.g., $E_{x}$$>$2 MeV) in combination with more precise knowledge of sharp resonance parameters $E_{\alpha}$=1.05 and 1.59 MeV of the $^{13}$C($\alpha$,$n$)$^{16}$O reaction.
 
\subsection{Target characterisation} \label{subsec:target}

For $^{13}$C, the application of solid-state targets is dominant in literature (see Table \ref{tab:exp_data_summary}). Carbon targets prepared from heating chemical compounds \cite{Bair_Haas1973, Ramström1976, Davids1968} or evaporated onto solid backing such as Ta or Cu \cite{Ciani2021, Heil2008, Brune1993, Drotleff1993, Harissopulos2005, Sekharan1967, deBoer2024} have been used. In the following, we review its various aspects.

\subsubsection{Methods of target characterisation} \label{subsubsec:target_charach}

Nuclear astrophysics experiments often require long exposures of target under intense beams, which can easily cause significant target degradation. Target modification processes (such as diffusion, melting, sputtering or contamination of the target surface \cite{PAINE1985, RAdek2017}) that occur under intense beam irradiation may result in significant changes of target stoichiometry as a function of depth and accumulated charge \cite{Wang1984}. There are different non-invasive ways to obtain information about target degradation in situ, e.g., measurements of the experimental yield variation as a function of time at a reference beam energy, Nuclear Resonant Reaction Analysis (NRRA), and the $\gamma$-shape technique. These methods are discussed in the following.

\begin{figure}[htb!]
    \centering
    \includegraphics[width=\columnwidth]{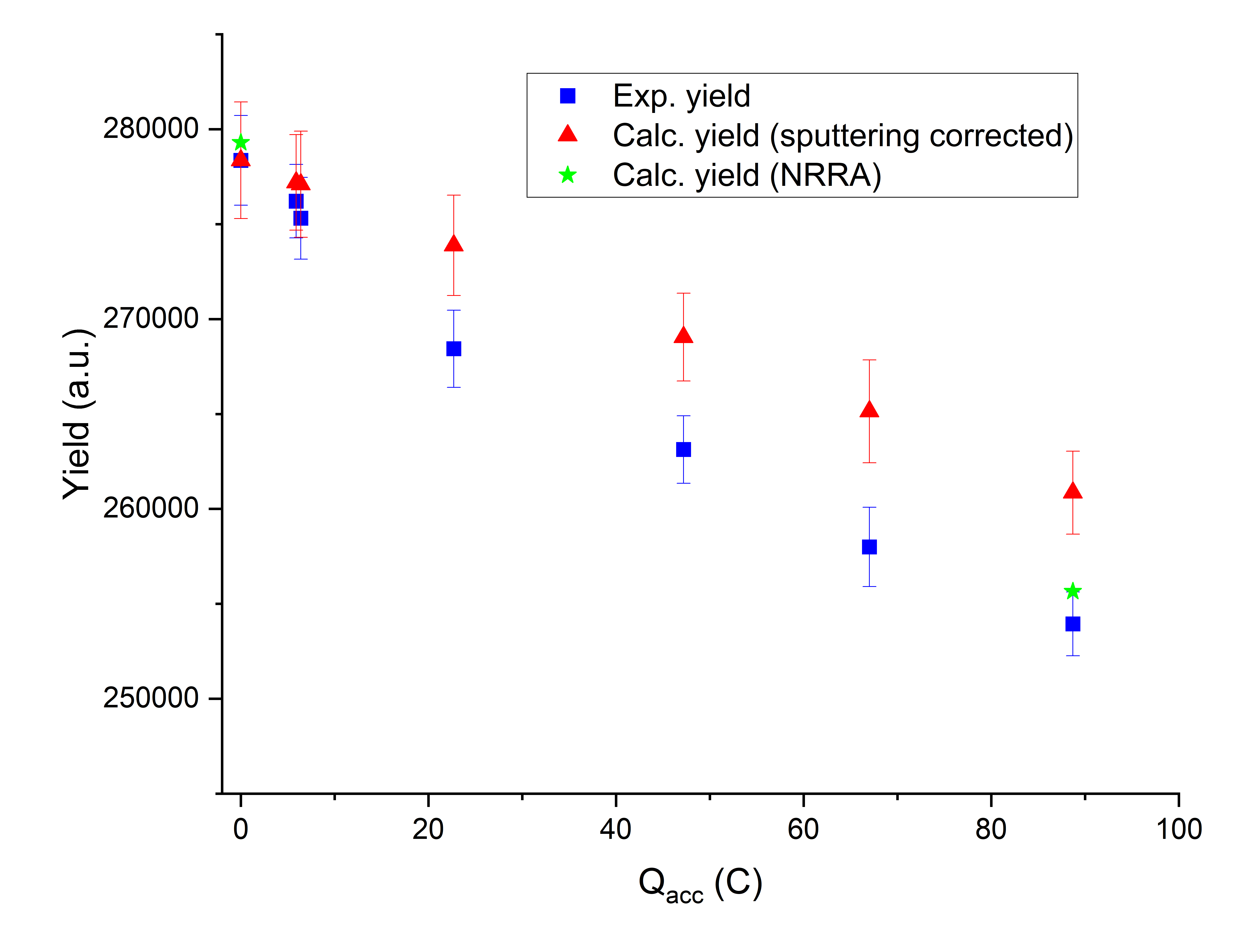}
    \caption{(Colour online) Reaction yield of $^{13}$C(p,$\gamma$)$^{14}$N as a function of accumulated charge. Blue squares represent the experimental yields, while red triangles show the calculated values based on cross section taken from \cite{KING1994} and corrected for sputtering effect from SRIM calculation. The green stars represent the calculated yields based on the NRRA scans.}
    \label{fig:Target_degrad}
\end{figure}

The overall effect of target degradation can be followed by the continuous monitoring of neutron or $\gamma$-ray yield (Y$_{\gamma}$) of a given transition in the investigated nuclear reaction \cite{Brune1993, Gao2022}. However, while target thinning caused by, e.g., sputtering can be well corrected in cross-section calculation from the time dependence of such a yield, other ion beam induced processes such as diffusion and mixing of the target atoms into/with the backing are more difficult to assess. 
The measured yields at LUNA400 of the $^{13}$C(p,$\gamma$)$^{14}$N reaction at E$_{p}$=0.31 MeV as a function of the accumulated charge on a single carbon target are shown in Figure~\ref{fig:Target_degrad} and represent the effect of modification through ion beam mixing. The calculated yields based on cross section taken from \cite{KING1994} if only target sputtering is considered are also plotted. The effect of this is estimated from a simulation using the SRIM software package \cite{SRIM}. It is evident that the sputtering effect alone can not explain the decreasing yield.
The calculated yields considering also the level of target modification obtained in NRRA method, are also plotted and agree well with the experimentally obtained yield. The NRRA method exploits the existence of a narrow and isolated resonance in a given reaction, whose cross section can be well described by the Breit-Wigner expression, $\sigma_{BW}$ \cite{Iliadis2007}. By measuring the yield as a function of beam energies in the proximity of the resonance on targets, a characteristic resonance yield curve can be obtained, which contains information on the target thickness and composition.

\begin{figure}[htb!]
    \centering
    \includegraphics[width=\columnwidth]{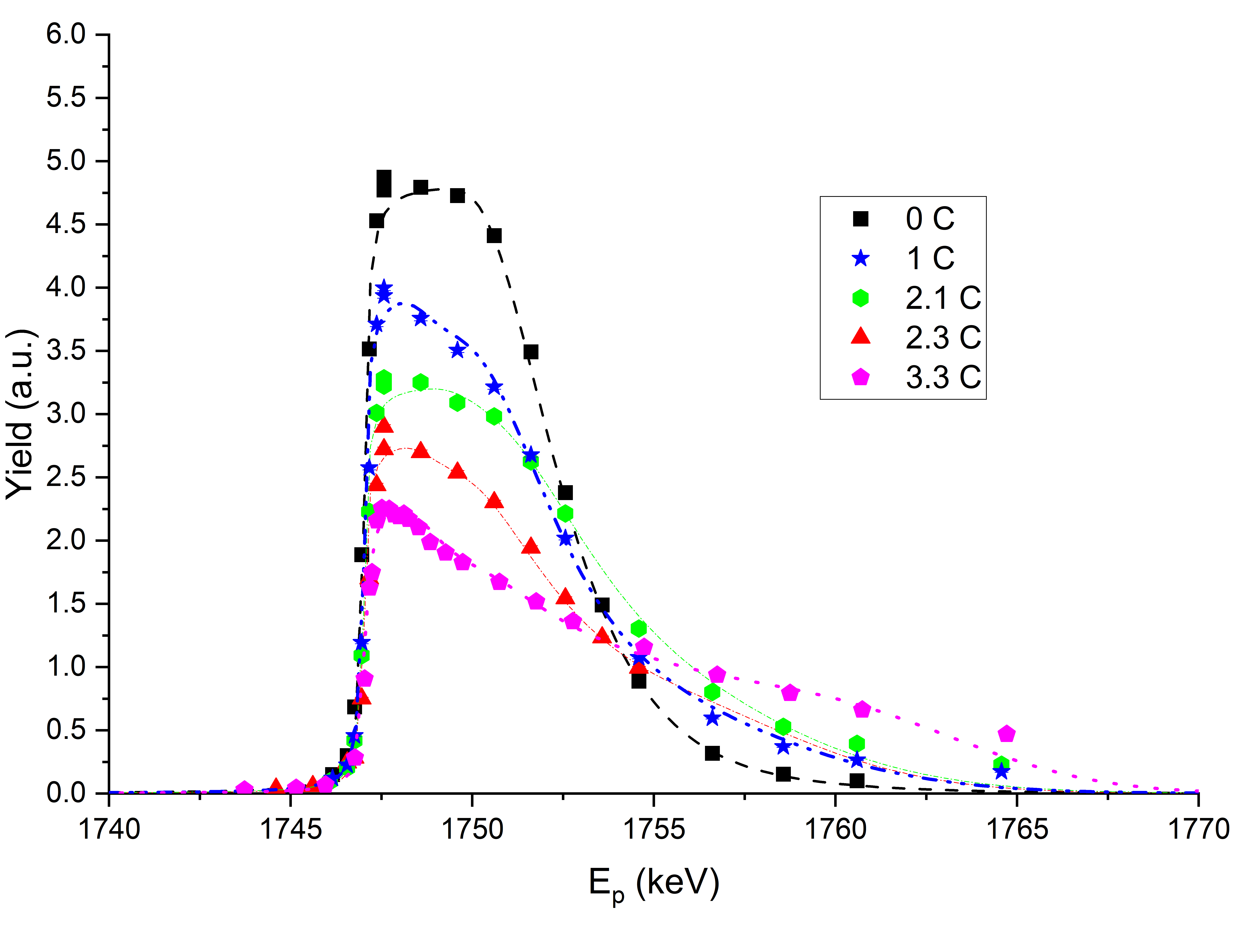}
    \caption{(Colour online) Resonance yield profiles measured on targets with different accumulated ($\alpha$-beam) charge. Lines are drawn to guide the eye.}
    \label{fig:Yield_profile}
\end{figure}

In the case of the LUNA experiment solid targets were produced by evaporating $^{13}$C enriched powder onto tantalum backings. Details of target preparation and characterisation are described in Ref.~\cite{Ciani2020}.
After the target production, a proton NRRA scan using the 1.747 MeV resonance of the $^{13}$C(p,$\gamma$)$^{14}$N reaction was performed to measure the initial target properties, such as number of active nuclei and target homogeneity. 
NRRA yield profile measurements at ATOMKI were repeated after different amounts of accumulated $\alpha$-beam charges (done at LUNA400). The result is shown in Figure~\ref{fig:Yield_profile}, where the appearance of beam induced target degradation is evident.

The observed modification of the measured target yield profile can be explained by the intense diffusion between the active target layer and the target backing, even without the inclusion of the light element implantation effect. 
This information should be used to correct low-energy cross-section measurements. It is also worth noting that repeated measurements of $^{13}$C($\alpha$,$n$)$^{16}$O experimental yield at higher E$_{\alpha}$ can be used to estimate target degradation \cite{Heil2008}, but this propagates uncertainty to the cross-section calculation at low E$_{\alpha}$, where the contribution of the surface layers to the experimental yield is more dominant, due to the exponential drop of the reaction cross section.
Thus, an alternative, in-situ technique to monitor target modification is often required. 

If no resonance is accessible, an alternative approach relies on the study of the shape of the $\gamma$-ray lines emitted in a suitable radiative capture reaction, for example in $^{13}$C(p,$\gamma$)$^{14}$N, to periodically check both the thickness and stoichiometry of the target. 
This so-called $\gamma$-shape analysis method was applied during the $^{13}$C($\alpha$,$n$)$^{16}$O campaign at LUNA \cite{Ciani2020}. To monitor the target degradation during the $^{13}$C($\alpha$,$n$)$^{16}$O measurements, data taking at LUNA400 consisted of long $\alpha$-beam runs with accumulated charges of $\sim$ 1 C per run, interspersed by short proton-beam runs with typical accumulated charges of 0.2 C at most, so as to minimize possible changes in target stoichiometry caused by the proton irradiation itself.
The maximum accumulated charge with $\alpha$ beam on each target was limited to 3\,C, corresponding to at most a 30\% target degradation due to the modification of stoichiometry.
The lowest studied energies of E$_{\alpha}$=0.245 and E$_{\alpha}$=0.233 \,MeV required special attention, as the statistics collected during a single run ($\sim$1 count/target) was insufficient to obtain a reliable estimate of the cross section. Thus, all runs at the same energy with similar target degradation level (obtained from $\gamma$-shape analysis) were summed up and the cross sections for each subset were calculated to obtain the cross section for the given energy \cite{Ciani_supplement}. 

\subsubsection{Stoichiometry of carbon target}  \label{subsubsec:Stoichiometry}

The $^{13}$C target stoichiometry was studied with different approaches.
The NRRA technique was used to measure the isotopic abundance of enriched $^{13}$C material based on the comparison of the plateau height of enriched and natural carbon targets with nominal $^{13}$C content of 99\% and 1.1\%, respectively. 
The $^{13}$C abundance of the powder used for LUNA was measured to be $(97.1\pm2.3)\%$, compatible with the 99\% reported by the manufacturer.

Another approach that can be used even at low projectile energies is based on the measured $\gamma$-ray intensities of DC$\rightarrow$g.s. of the $^{12,13}$C(p,$\gamma$)$^{13,14}$N reactions. A target was irradiated at E$_{p}$=0.31 MeV at LUNA400 and the E$_{\gamma}$=2.23 MeV (from $^{12}$C(p,$\gamma$)$^{13}$N) and 7.84 MeV (from $^{13}$C(p,$\gamma$)$^{14}$N) lines were detected using a HPGe detector. Based on the interpolated cross-section data from \cite{KING1994} and \cite{Azuma2010} and on the efficiency measurement with radioactive sources and radiative capture reactions, the isotopic abundance of the $^{13}$C target can be extracted. In the LUNA experiment, (98$\pm$2)\% was obtained, which agrees well with the data obtained from the NRRA measurement and with the value indicated by the manufacturer.

\subsubsection{Light element contamination} \label{subsubsec:light_element}

The presence of contaminants and their time dependent accumulation in the irradiated targets requires a special attention. At low energy, deuterium, boron isotopes, carbon, oxygen and fluorine are considered the main contributors of beam-induced background in proton-induced reactions \cite{Best2016_2}, while in $\alpha$-induced reactions the $^9$Be, boron isotopes (especially $^{11}$B through the $^{11}$B($\alpha$,$n$)$^{14}$N reaction \cite{deBoer_2020_11B}), $^{13}$C and $^{17}$O might contribute to the neutron background of the experimental apparatus. 
To prevent carbon built-up on target surface during the irradiation good vacuum conditions often supplemented with liquid nitrogen cooled cooling traps are necessary. The effect of possible carbon build-up, e.g., are shown in Figure 4 of Ref. \cite{Heil2008} in the case of a thin-target and it is also considered as a potential source of uncertainty in \cite{Gao2022} in thick target arrangements.
In the following we present the different approaches used in the LUNA experiment to quantify the presence of light contaminants and their origin.

The level of boron contamination in the target was measured by Particle Induced Gamma-ray Emission (PIGE) technique at ATOMKI at $E_p$=3.23 MeV \footnote{The proton energy was chosen considering the neutron threshold of the $^{13}$C(p,n)$^{13}$N reaction at $E_{p}$=3235.55$\pm$0.29 keV.}. Boron content of the target was calculated based on the experimental thick-target yield \cite{Chiari2016} as well as on thin target data using the well-known cross section of the $^{11}$B($\alpha$,$n$)$^{14}$N reaction \cite{Preketes2016}. An upper limit of $<$ 6 ppm corresponding to about $<$ 4x10$^{14}$ atom/cm$^{2}$ was obtained. 

To further investigate the possible contribution of beam induced background in the LUNA $^{13}$C($\alpha$,$n$)$^{16}$O experiment, an evaporated $^{13}$C target was irradiated with proton beam at $E_{p}$=0.31 MeV with a total 6.7 C accumulated charge. This was compared to a beam induced background study on a blank target (Ta backing) with a total 2.7 C accumulated charge at $E_{p}$=0.38 MeV.
Observation were in line with the results obtained with PIGE and no evidence of, e.g., boron contaminants was observed in the gamma-ray spectrum.

To further check the effect of possible light contaminants on $\epsilon_{\rm{eff}}$(E), calculations using SRIM for proton energies at $E_{p}$ = 0.28$-$0.31 MeV, and energies at $E_{\alpha}$ =0.3$-$0.4 MeV relevant to the $^{13}$C($\alpha$,$n$)$^{16}$O data taking campaign were performed. Assuming the presence of only one light element contaminant in the target (H, He, C or O), $\epsilon_{\rm{eff}}$(E) changes by less than 3\% for proton projectiles and less than 5\% for $\alpha$ particles in the projectile energy ranges used in the simulation. This can be also concluded if more than one contaminant is present at the same time. 
According to this observations, the $\gamma$-shape technique can quantify the stopping power modification independently from the type of contaminant \cite{Ciani_PhD}.

The low level of contaminant in LUNA targets were further supported by additional measurements using Elastic Recoil Detection Analysis (ERDA) performed on irradiated targets at the Ion Beam Center of Helmholtz-Zentrum Dresden-Rossendorf \cite{ERDA}.
The analysis confirmed that the concentration of elements such as H, He and O after the $\alpha$-beam irradiation at LUNA was at most 10\%. 

Similarly to the LUNA experiment, in Refs. \cite{Heil2008, Drotleff1993, Brune1993, Harissopulos2005, Ciani2021, deBoer2024} isotopically enriched $^{13}$C targets were evaporated onto heavy elements backings (e.g., Cu, Ta, Au) and the above indicated target degradation effect was controlled either with NRRA method or with yield measurements at a reference $\alpha$ energy.
Refs. \cite{Gao2022, Davids1968, Bair_Haas1973, Ramström1976} used thick targets in their experiments. Thick target measurements could be suited to avoid systematic uncertainty, however, in these cases light particle implantation requires a special attention, discussed in the next section.

\subsubsection{Effect of high dose $\alpha$ implantation on the experimental yield} \label{subsubsec:implant}

In the case of low energy nuclear astrophysics experiments, the number of accumulated beam particles may become comparable with the number of active nuclei of thin targets. For example assuming 100 $\mu$A beam intensity and 1 hour irradiation time, even $\sim$6x10$^{18}$ atom/cm$^{2}$ implantation dose is reached with a typical size of a few mm lateral beam size. 
Therefore, the alteration of the stoichiometry of the target due to the implantation itself 
and thus the modification of effective stopping power may increase drastically the uncertainty of the calculated cross section.The implantation effect should also be considered in thick target measurement, especially when a single target is used in a long irradiation campaign \cite{Gao2022, Harissopulos2005}.

\begin{figure}[htb!]
    \centering
    \includegraphics[width=\columnwidth]{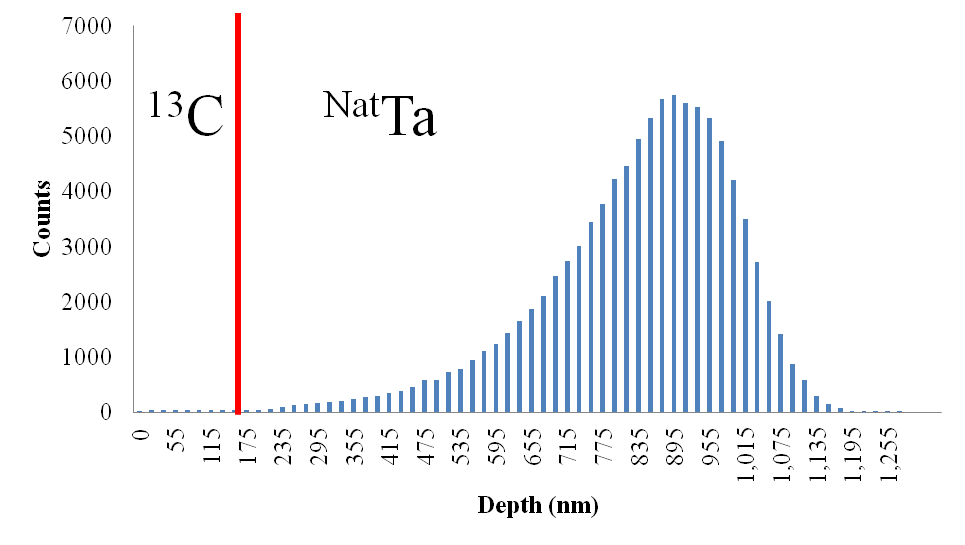}
    \caption{(Colour online) Simulation of $E_{\alpha}$=0.4 MeV alpha particle implantation in carbon target evaporated onto thick Ta backing.}
    \label{fig:TRIM_calc}
\end{figure}

Simulation of the implantation using $E_{\alpha}$=0.4 MeV \footnote{In line with the LUNA experiment} alpha particles impinging on a thin evaporated $^{13}$C target is shown in Figure ~\ref{fig:TRIM_calc} using SRIM software. 

Although the simulation reveals that the bombarding particles are implanted in a finite depth range in the Ta backings, well-separated from the evaporated thin layer, they might still affect the cross-section calculation using a thick target over a wide alpha-energy range, e.g., using repeated yield measurement at high $E_{\alpha}$ for target monitoring at low $E_{\alpha}$ run \cite{Heil2008, Gao2022}.

\begin{figure}[htb!]
    \centering
    \includegraphics[width=\columnwidth]{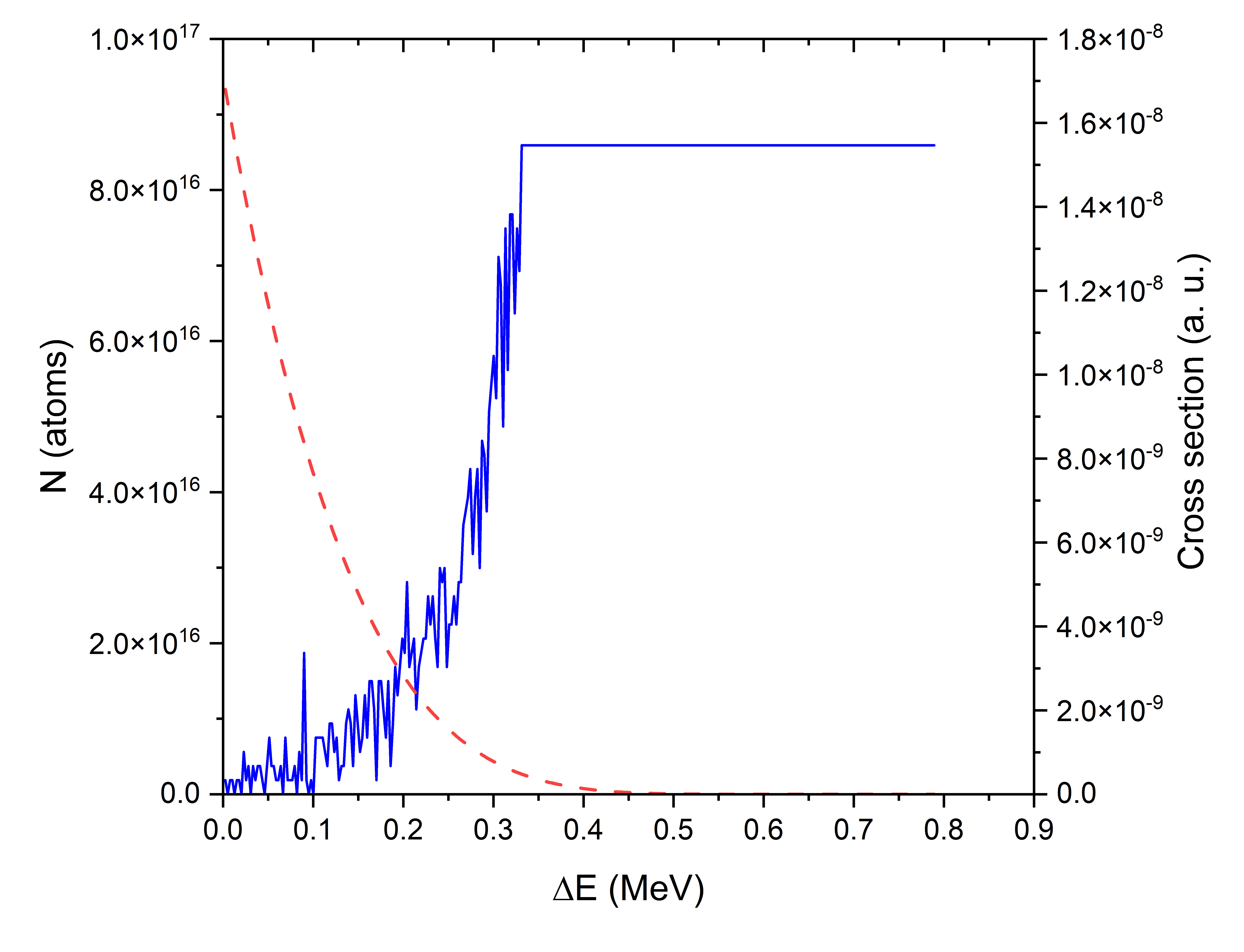}
    \caption{(Colour online) Calculated cross section of $^{13}$C($\alpha$,$n$)$^{16}$O reaction (dashed red line) and helium profile (solid blue line) in thick carbon target at $E_{\alpha}$=0.8 MeV beam after irradiation of $E_{\alpha}$=0.4 MeV with a total of 3 C accumulated charge.}
    \label{fig:SRIM_He_Graf}
\end{figure}

To estimate the effect of light nuclei implantation into carbon on the measured neutron yield in the case of a thick carbon target, another simulation was carried out using the SRIM software package. First, the target structure was modelled assuming 3 C implantation at $E_{\alpha}$=0.4 MeV. Then, the integrated yield at a $E_{\alpha}$=0.8 MeV\footnote{This energy was selected in line with \cite{Heil2008}} was recalculated and compared with the non-irradiated case.
An important parameter of the simulation is the saturation level of a given light nucleus in a heavier matrix. Although irradiation of carbon with light nuclei ($^{1}$H, $^{4}$He) has a significant interest in nuclear technology and material science \cite{ATSUMI1986, Moller1982, Alimov1995}, limited experimental information is available in literature with intense $^{+}$He beam in the energy region of nuclear astrophysics interest. Wide range of indicated saturation level are suggested \cite{Alimov1995, TERREAULT1980}. In our simulation, 40\% saturation level \cite{TERREAULT1980} was used and the results of the calculation is shown in Figure ~\ref{fig:SRIM_He_Graf}. 

It is assumed that helium atoms above the saturation limit diffuse towards deeper layers of the target, which agree with the experimentally observed yield reduction in proton runs. If different diffusion directions (e.g., towards the surface of the target) are assumed, the predicted yield reduction disagrees with observations. Since at 3 C of accumulated charge the saturation limit is reached, adding more charge does not change the structure of layers near the surface, which mainly contribute to the observed neutron yields. 

In summary, $\sim$ 1.6\% deviation of the experimental yields compared to the non-irradiated case is obtained assuming 40\% saturation level, which reveals the marginal role of implantation effect in cross-section calculation.

\section{Summary of currently available direct measurement data} \label{sec:overview}

Despite many attempts to measure the $^{13}$C($\alpha$,$n$)$^{16}$O cross section \cite{Davids1968,Bair_Haas1973,Kellog1989,Drotleff1993,Brune1993,Harissopulos2005,Heil2008}, direct measurements inside the s-process Gamow window $E_{\rm{c.m.}}$=0.14$-$0.25 MeV have been performed only recently \cite{Ciani2021,Gao2022} and covering only its highest energy part. Thus, the theoretical approach to extrapolate the $S(E)$-factor in the astrophysically relevant energy region is still mandatory. 
Extrapolation based on experimental cross-section data is complicated by the different systematic uncertainties of the different studies and because in many cases the full description of uncertainty budget is not available.
The uncertainties of the available experimental data of the $^{13}$C($\alpha$,$n$)$^{16}$O reaction are summarised in Table \ref{tab:resonance_param}.

\begin{table}[t!]
\caption{Uncertainty budget of direct cross-section measurements of the $^{13}$C($\alpha$,$n$)$^{16}$O reaction. The tabulated values are given in \% unless noted otherwise.}
\begin{center}
\resizebox{1\columnwidth}{!}{
\begin{tabular}{|c | c | c | c | c | c | c |}
\hline
Data & Stat. & \multicolumn{4}{c}{Sys.} & E. calib. \\ \hline

 & & $\eta_{n}$ & Target thickness & Stopping power & Beam current  & \\ \hline

Heil et al. \cite{Heil2008} & 0.1-92 & \multicolumn{4}{c}{5-40} & n.d.  \\

Bair et al. \cite{Bair_Haas1973} & n.d. & \multicolumn{4}{c}{15-18} & 0.15\%\\

Davids  \cite{Davids1968} & 2.3-10.7 & 10$^{a}$ & n.d. & 10 & 5 & \\

Drotleff et al. \cite{Drotleff1993} & n.d. & \multicolumn{5}{c}{n.d.} \\

Brune et al. \cite{Brune1993} & & & & & & 0.1\% \\

Harrisopulos et al. \cite{Harissopulos2005} & 1.6$^{a}$ & 1.7 & 3 & 3 &  2 &  3 keV \\

Ramström and Wiedling  \cite{Ramström1976}& & 8.2 & & & & \\

Kellogg et al. \cite{Kellog1989}& & & & & & \\

Sekharan et al. \cite{Sekharan1967}& & 12 & \multicolumn{2}{c}{16}  & 2 & 3 keV \\

Ciani et al. \cite{Ciani2021}& 2.2-18.1 & 8 &   & 5 & 3 & 0.1\% \\

Gao et al.$^{b}$ \cite{Gao2022}& $<$15; 3-8$^{c}$; 2-3$^{d}$& 7 &  & 6 & 5 & n.d. \\

Febbraro et al. \cite{Febbraro2020}& \multicolumn{5}{c}{Origin of uncertainties not described} \\

Brandenburg et al.$^{e}$  \cite{Brandenburg2023} & 1-7 & 16 & 4 & 7 & 1 & 0.2\% \\

deBoer \cite{deBoer2024} & $<$10 & 5 and 10 $^{f}$ & 5 & 3 & \\

 \hline
\end{tabular}
\label{tab:resonance_param}
}
\end{center}
\footnotesize $a$ At E$_{\alpha}$= 1 MeV.\\
\footnotesize $b$ 2 and 4\% implied by angular distribution.\\
\footnotesize $c$ Originated from beam tuning and possible carbon build-up.\\
\footnotesize $d$ Reproducibility of thin and thick target measurement.\\
\footnotesize $e$ 10 and 14\% implied by angular distribution in $<$5 MeV and $>$5 MeV, respectively.\\
\footnotesize $f$ intristic and MCNP/geometry efficiency, respectively.\\
\end{table}

To solve the inconsistency of the different experimental data-sets, normalization factors of experimental data can be applied. Normalization factors as high as 40\% have been reported by various authors, which are well beyond the reported systematic uncertainties of the experimental data, usually less than 20\%. 

For example, while Drotleff et al. \cite{Drotleff1993} did not mention this normalization issue, some inconsistencies between data sets were already noted in previous works (see, e.g., Bair \& Haas \cite{Bair_Haas1973} and Brune \cite{Brune1993}). 
Harissopulos et al. \cite{Harissopulos2005} re-scaled their data averaging their cross section at $E_{\alpha}$ = 1.00 MeV and compared it with that of Brune et al. \cite{Brune1993}, the difference was $<$10\%, inside the uncertainty bands. Moreover, their values were found to be lower by 28-37$\%$ with respect to Bair \& Haas \cite{Bair_Haas1973}, instead Heil et al. \cite{Heil2008} presented a good agreement with Davids \cite{Davids1968}, while Bair \& Haas \cite{Bair_Haas1973} and Drotleff et al. \cite{Drotleff1993} have a poorer agreement (15$\%$ discrepancy) with Kellogg \cite{Kellog1989} and Harissopulos et. al  \cite{Harissopulos2005}. 

Correction factors of 1.36 and 0.7 are proposed by Giorginis et al. \cite{Giorginis2016} and Plompen et al. \cite{Plompen2013} be applied to the data of Harissopulos et al. \cite{Harissopulos2005} and Bair \& Haas \cite{Bair_Haas1973}, respectively, which may originate from the inaccuracy of target thickness determination and neutron detection efficiency calculation.
Based on the analysis of Ref. \cite{Giorginis2016}, Ref. \cite{CHADWICK2018} reported that a correction of 1.36 has to be applied to the number of active nuclei due to the incorrect extraction of target thickness from the resonance profile obtained by Harissopulos et al. \cite{Harissopulos2005} at $E_{p}$=1747 keV in $^{13}$C(p,$\gamma$)$^{14}$N reaction. They mention that the comparable value of target thickness and energy resolution function requires a deconvolution (via Voight profile) analysis, however, they did not discuss the target profile obtained by Harissopulos et al. \cite{Harissopulos2005} at $E_{\alpha}$= 1054 and 1336 keV, which is in agreement with the target thickness at $E_{p}$=1747 keV and therefore does not support the above proposed correction factor.

Bair \& Haas \cite{Bair_Haas1973} used age-diffusion theory calculation to obtain neutron detection efficiency. Details of that work can be found in Macklin et al. \cite{MACKLIN1957}. In this context, Ref. \cite{Plompen2013} refers to the need of different corrections due to geometrical (e.g., tube pressure) and angular distribution effects recommending the reevaluation of this data.\\

Recent cross-section measurements will be discussed in the context of low energy extrapolation in the following section.

\section{R-matrix approach for low-energy extrapolations} \label{sec:R_matrix}

In spite of the available cross-section data, low energy extrapolations are still required by stellar models. R-Matrix theory has been proven to be a reliable tool for such extrapolations \cite{descouvemont2010}. LUNA implemented R-Matrix through the code AZURE2 \cite{Azuma2010} and included the resonance parameters proposed by  \cite{Tilley1993}, \cite{Avila2015} and \cite{leal2016} and reported in \cite{deBoer2020}. Monte Carlo fits were performed by fixing some parameters from literature and fitting others, as described in the following. The cross-section data included in this analysis were those from Heil et al. \cite{Heil2008}, Drotleff et al. \cite{Drotleff1993} and Harissopulos et al. \cite{Harissopulos2005}. 

The LUNA analysis was limited up to $\rm E_{cm}$ = 1.2 MeV, a range which includes the whole Gamow window for temperature $\rm \leq1~GK$. 
Two broad resonances are relevant in this range: one located at $\rm E_{cm}$ = 856 keV and the sub-threshold one at $\rm E_{cm}$ = $-$2.7 keV.
The resonance parameter of the latter was used from indirect investigation (e.g., \cite{Tilley1993}) as no close enough direct experimental data exist. 
Resonance parameters were allowed to vary in the calculation within their reported uncertainties, i.e., ANC= $\rm 5.44~\times~10^{90}~fm^{-1/2}$ (19$\%$) from \cite{Avila2015} (as also confirmed by \cite{Trippella&LaCognata2017}) and $\Gamma_{n}$ = 124 keV (10$\%$) from \cite{Tilley1993}. The 856 keV resonance energy was fixed, being its influence negligible in the reaction rate evaluation. For each spin, a background pole at $E_{cm}$=15 MeV  was included to represent the non-resonant component of the cross section (e.g., the contribution from low energy tail of higher energy resonances). Based on the trends and absolute scale of experimental data, results of \cite{Harissopulos2005, Kellog1989} and of \cite{Gao2022, Bair_Haas1973, Heil2008, Drotleff1993} form two well-separated subgroups.

\begin{figure*}[ht!]
\begin{center}
    \begin{tabular}{c c}
    \includegraphics[trim={0.45cm 0 .95cm 0},clip,width=1\columnwidth]{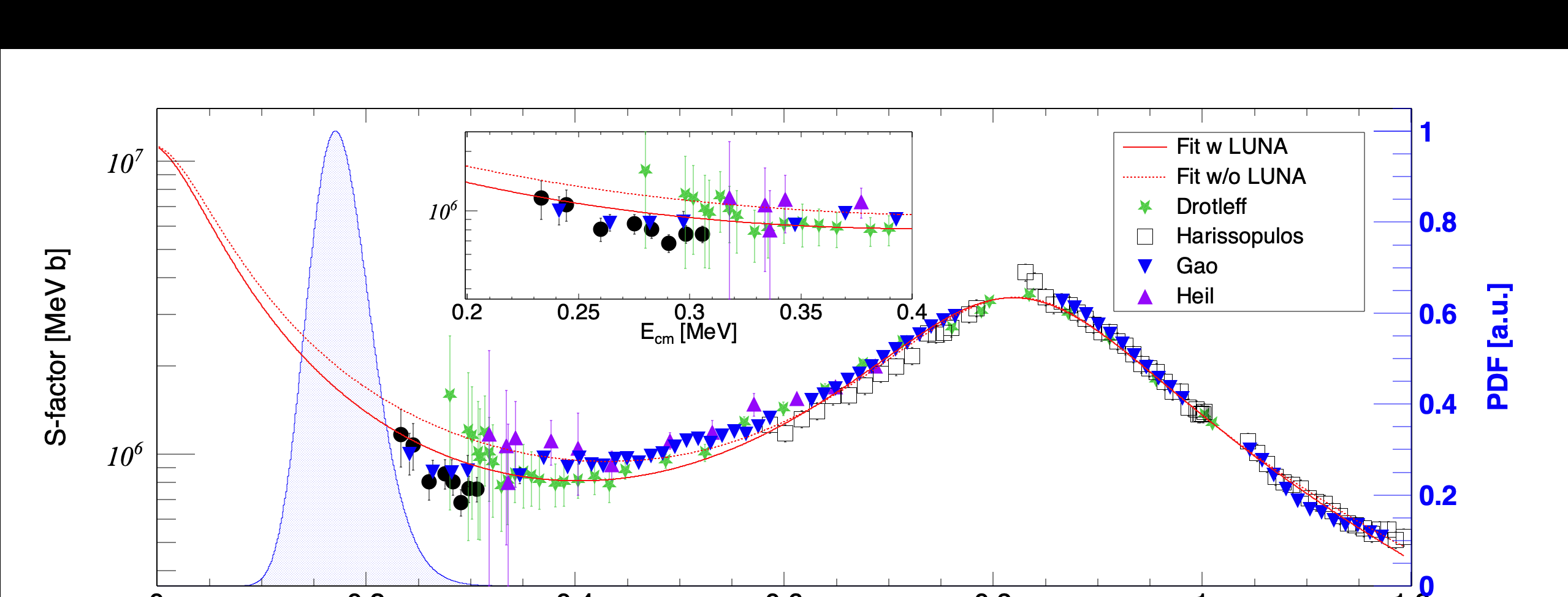}
    &
    \includegraphics[trim={0.45cm 0 .95cm 0},clip,width=1\columnwidth]{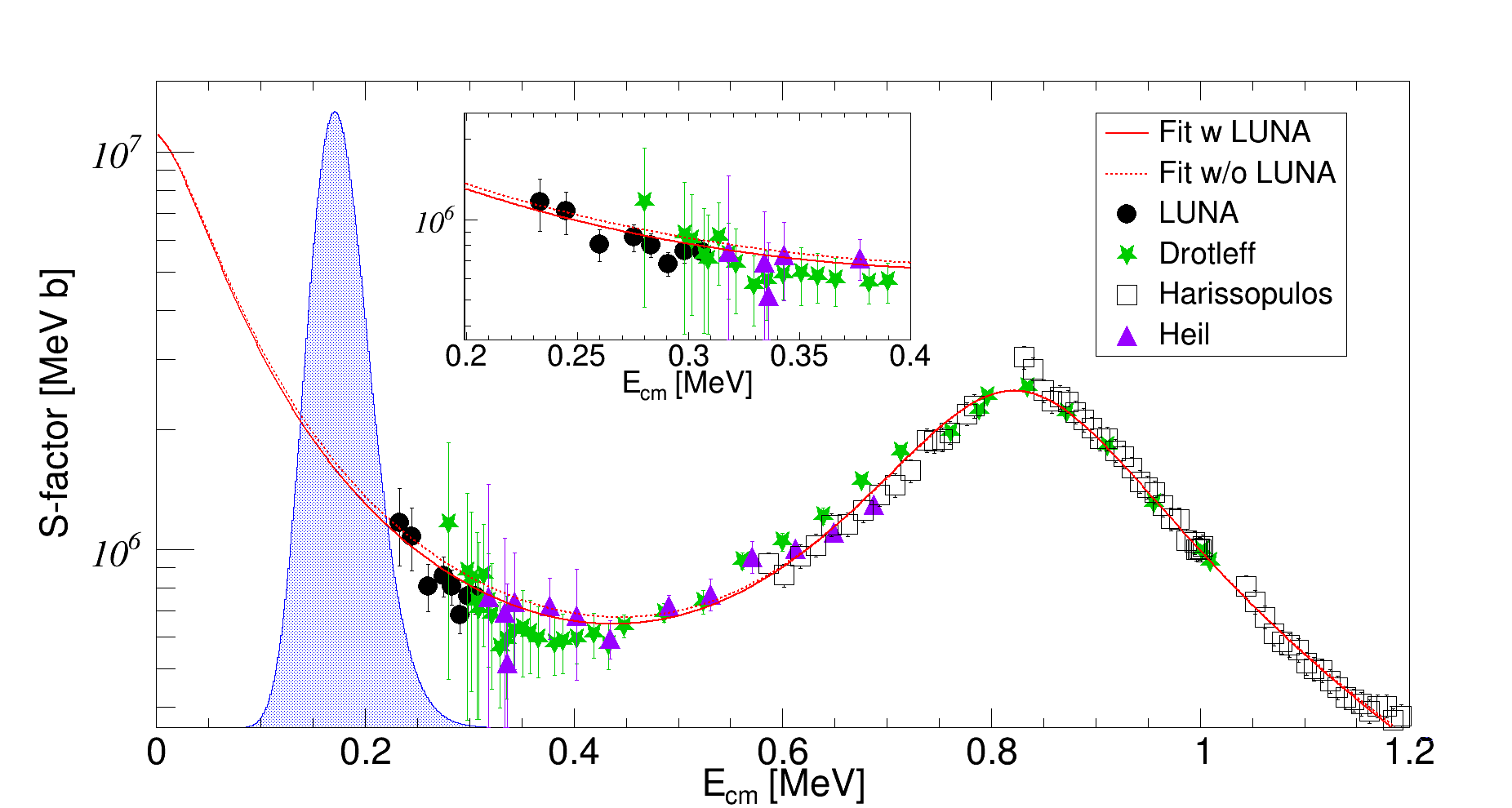}
    \end{tabular}
\end{center}
\caption{(Colour online) Left: Astrophysical $S(E)$-factor of $^{13}$C$(\alpha$,n$)^{16}$O using \cite{Heil2008} and \cite{Drotleff1993} normalization reference. The lines show the results of two R-matrix analyses, with and without LUNA data \cite{Ciani2021}.  
The fit from ref. \cite{Ciani2021} was updated adding results from \cite{deBoer2024} and \cite{Gao2022} with their original normalizations.
Right: Fit performed with \cite{Harissopulos2005} as normalization reference. Data of Ref. \cite{Gao2022} and Ref. \cite{deBoer2024} was scaled by a factor 0.73. Figure adapted from \cite{rapagnani2023}.}
\label{fig:sfactor}
\end{figure*}

The effect of the normalisation factor of the experimental data \cite{Ciani2021} was investigated via a dedicated analysis.
First, the data of Ref. \cite{Heil2008} and \cite{Drotleff1993} were considered as reference and a normalization factor for Ref. \cite{Harissopulos2005} data was included to be fit together with the resonance parameters. Second, Ref. \cite{Harissopulos2005} data was considered as a reference and a normalization factor was applied to Ref. \cite{Heil2008} and \cite{Drotleff1993}, again to be fitted with resonance parameters. For the LUNA data, considering the description of possible systematic uncertainties, no normalization factor was applied. 
Results are summarized in Figure \ref{fig:sfactor}. In the first case, the fitted normalization factor of 1.37(16), very close to 1.36 proposed by \cite{CHADWICK2018} was obtained (Left side of the Figure), while the value of 0.73(9) was used in the second case (Right side of the Figure). 
Inside the Gamow peak of the s-process, the effect of the different normalisation factor is only of the order of 5\% over the extrapolated $S(E)$-factor, increasing toward higher energies.

To estimate the $S(E)$-factor uncertainty, 30'000 evaluations for each normalization group were performed with AZURE2 sampling input parameters with a Gaussian probability within their experimental uncertainty as well as sub-threshold state ANC. Poles and the partial width of 856 keV resonance were fit to data. The resulting $S(E)$-factors, extrapolated down to zero energy, were used to build a probability density function (PDF) from which the reaction rate was evaluated at several temperatures up to 1 GK, as reported in \cite{Ciani2021, Ciani_supplement}. Considering the smaller tension with the data from Ref. \cite{Heil2008} and \cite{Drotleff1993}, the normalization factor of 1.37(16) for \cite{Harissopulos2005} was used in the evaluation of best reaction rates reported in Ciani et al. \cite{Ciani2021}.
The "low-LUNA" rates, which are the present lower limit for the reaction rate, were obtained by extracting the 5\% percentile line of the $S(E)$-factor PDF evaluated from the normalization of \cite{Heil2008} and \cite{Drotleff1993} data with 0.73(9) (based on Ref.\cite{Harissopulos2005}).
Considering that LUNA data are very close to the s-process Gamow window, fits are well constrained in such energy range in spite of the normalization used. At higher energies at interest for the i-process, the ambiguity on data normalization is still relevant and deserves a deeper investigation. 

Since the publication of the LUNA result, new directly measured cross-section data have been published in Ref. \cite{Gao2022} and Ref. \cite{deBoer2024}. 

JUNA collaboration \cite{Gao2022} provides a consistent data set from 0.24 to 1.9 MeV obtained in two experimental campaigns at two accelerator facilities. Even if the JUNA data of different campaigns do not overlap in energy, they seem to agree with the normalisation of Ref. \cite{Heil2008} and \cite{Drotleff1993} (data are also shown in Figure \ref{fig:sfactor}). This makes the LUNA assumption for the recommended value more robust in the whole considered temperature range. It is worth noting that JUNA performed R-matrix fit excluding other literature data to obtain ANC of sub-threshold state and extrapolated $S(E)$-factor at $E_{\rm{c.m.}}$=0.24 MeV and below. Their Coulomb renormalized ANC value $\overset{\sim}{\rm C}^{2}$=2.1(5) fm$^{-1}$ is in tension with $\overset{\sim}{\rm C}^{2}$=3.6(7) fm$^{-1}$ of Ref. \cite{Avila2015}, which was used in the LUNA analysis. The different ANC values imply uncertainty of extrapolation mainly below E$_{\rm{c.m.}}$=0.24 MeV (see Figure 2. of \cite{Gao2022}), where no direct data exist. 

Recently, a new value for the ANC $\overset{\sim}{\rm C}^{2}$=2.8(5) fm$^{-1}$ has been proposed \cite{hebborn2023}, which decreases the estimate of \cite{Avila2015} about 22\%. We performed an additional analysis including this value (used Refs. \cite{Heil2008,Drotleff1993} as normalization reference) and found a difference of 15\% in the reaction rate at 90 MK. If this new ANC is confirmed, a deeper investigation of its effects on reaction rates would be desirable.

Ref. \cite{deBoer2024} published high resolution differential cross-section data in $E_{\rm{c.m.}}$=0.6-5.0 MeV. Their result, similarly to Ref. \cite{Gao2022}, favours the absolute cross section published in Refs. \cite{Heil2008, Drotleff1993}. Their angle integrated cross section obtained from Legedre fit of differential cross-section data are also shown in Figure \ref{fig:sfactor}. These authors performed a detailed R-matrix analysis to obtain cross section over the Gamow energy range. First, an R-matrix analysis completed with Bayesian uncertainty analysis was done using the experimental data of Refs. \cite{Heil2008, Drotleff1993, Ciani2021, Gao2022} and $^{16}$O(n, total) data. The obtained best fit suggests a $\approx$10\% uncertainty over the Gamow energy range. Then, the analysis was repeated using their differential cross-section data. The two fits are consistent, but the latter results a reduced $\approx$5\% uncertainty of $S(E)$-factor in Gamow energy. Experimental data were rescaled by Ref. \cite{deBoer2024} in their analysis by a larger factor in the case of Ref. \cite{Ciani2021} than the quoted systematic uncertainty. Moreover, the two data set of JUNA work Ref. \cite{Gao2022} were handled independently to obtain their best fit. Overall, this analysis highlighted the importance of low-energy angle integrated cross-section measurement and the potential of their fit in Gamow energy range, they concluded that a full evaluation of all the literature data is still needed for the uncertainty analysis of the extrapolated $S(E)$-factors.

\begin{sidewaystable*}[ht!]

\caption{Selected experimental setups and parameters of direct measurements of the $^{13}$C($\alpha$,$n$)$^{16}$O reaction}
\begin{center}
\resizebox{1\columnwidth}{!}{
\begin{tabular}{|p{1.7cm} | p{2cm} | p{1.2cm} | p{3cm} | p{2.2cm} | p{3cm} | p{3cm} | p{2cm} | p{5cm} | p{1.5cm} |p{3cm} | p{2cm} |}
\hline

Reference	&	E$_{\alpha}$ (MeV) & Int. ($\mu$A) & Accelerator & Accelerator Calibration & Detector & Detector Calibration & Solid angle & Target & Thickness & Method & Laboratory Background \\ \hline

Sekharan \cite{Sekharan1967} & 1.95-5.57 & & Bombay, VdG & & BF$_{3}$ based counters embedded in paraffin & $^{7}$Li(p,n)$^{7}$Be , Ra-$\alpha$-Be & 4$\pi$ & Electrically enriched $^{13}$C (30\%) onto 0.25 mm Ta & & &  \\

Davids \cite{Davids1968} & 0.475-0.7 & 25- 30 & Kellogg Radiation Laboratory, Oak Ridge type RF ion source & & Stilbene crystal and PSD & Not clear & 0° & Thick $^{13}$C target from CH$_{3}$I enriched to 54\% in $^{13}$C , Total 1.3 C was collected & 200 $\mu$g/cm$^{2}$ & $^{13}$C(p,$\gamma$)$^{14}$N resonance, but E$_{r}$ is not indicated & \\

Bair \cite{Bair_Haas1973} & 1-5  & - & Oak Ridge NAtional Lab. 5.5 MV Van de Graaff & $^{19}$F($\alpha$,$n$)$^{22}$Na, $^{7}$Li(p,n)$^{11}$B, $^{18}$O($\alpha$,$n$)$^{21}$Ne & Graphite-sphere neutron detector, 8 -$^{10}$BF$_{3}$ detector & in \cite{MACKLIN1957} Age-diffusion Theory, recalib. with Standard Sources & 4 $\pi$ & Infinitely thick disk of compressed carbon enriched $^{13}$C and thin $^{13}$C target produced by cracking enriched acetylene onto Pt backing & 5 keV at 1 MeV & & 2-3 c/s based on \cite{MACKLIN1957} \\

Ramström \cite{Ramström1976} & 0.6-1.15 & & Neutron Physics Laboratory, Nyköping, Sweden, 5.5 MV Van de Graaff & E$_{\alpha}$=1.05 MeV was repeated, 5 keV was observed and used in corrections & 20 $^{10}$BF$_{3}$ 0.7 m shielding of paraffin, Cd and concrete & PuBe, RaBe, $^{51}$V(p,n)$^{51}$Cr (at 2.3 MeV) Assuming non-energy dependent detector function (18.3$\pm$1.5) & 4$\pi$ & Methyl Iodine heated onto Ta 89\% $^{13}$ & 88 and 13 (in 0.7 and 1 MeV) $\mu$g/cm$^{2}$ & Weighting and width of profile at E$_{\alpha}$=1.05 MeV & \\

Kellogg \cite{Kellog1989} & 0.4-1.2 & & Calltech pelletron & & Neutron detector with 23\% efficiency-no details & & 4$\pi$ & & & & 0.027 c/s \\

Drotleff \cite{Drotleff1993}$^{a}$ & 0.35-1.4 & 100 & Stuttgart 4 MV Dynamitron & & 2 concentric cycle of 8 $^{3}$He counter in PE plus layers of PE, Paraffin, B and Cd & $^{252}$Cf, E dependence was calculated with multiproup calculation & 4 $\pi$ & 99\% $^{13}$C on solid Cu backing & & & 0.08 c/s \\

Brune \cite{Brune1993}$^{b}$ & 0.45-1.05 & 50 & Calltech pelletron & & 11 $^{3}$He filled counter in PE \cite{Wang1991} & $^{252}$Cf (20.2\% efficiency obtained), \cite{Wang1991} $^{7}$Li(p,n) reaction eff. is constant 50 keV and 2 MeV within ~ 5\% & 4$\pi$ & Thin $^{13}$C target (99.2\%) onto Cu disk, electron beam evaporated & & From known $^{13}$C($\alpha$,$n$) yield of non-resonance range and resonance strength (refer to \cite{Brune1993} and \cite{Bair_Haas1973} & 0.1 c/s \\

Harissopulos \cite{Harissopulos2005} & 0.8-8 & 100 nA & Ruhr-Universitat, Bochum, Dynamitron-Tandem Laboratory & & 8 $^{3}$He counter at 16 cm,8 $^{3}$He counter at 24 cm, Embeded in PE passive shielding (Cd, PE, B-PE, B-parafin) & $^{252}$Cf with MCNP simulation , Above 6 MeV n1, n2 branching & 4 $\pi$ & 99\% $^{13}$C on Ta backing. Air cooled target 40 mm, target degrad. is partially explained. Yield test at selected energies gave 2\% reproducibility. Presumably 1 target was used in this measurement. & 22 $\mu$g/cm$^{2}$, ~1e$^{18}$ atom/cm$^{2}$ & NRRA using $^{13}$C(p,$\gamma$)$^{14}$N at E$_{p}$=1.75 MeV & 0.22 c/s \\

Heil \cite{Heil2008} & 0.416-0.899 & 50 & Karlsruhe 3.7 MV van de Graaff & $^{7}$Li(p,n)$^{11}$B, E$_{r}$=402, 814, 953 keV & 42xBF$_{2}$ n/$\gamma$ converter using $^{113}$Cd(n, $\gamma$)$^{114}$Cd & $^{51}$V(p,n)$^{51}$Cr E$_{n}$=135, 935, 1935 keV GEANT4 simulation& 4 $\pi$ & $^{13}$C(99\%) electron gun onto 5 $\mu$m Au and Cu sheet, Impurities of Cu is not discussed & 7 keV at E$_{p}$=448.5 keV & NRRA $^{13}$C(p,$\gamma$)$^{14}$N at E$_{p}$=448.5 keV, yield check at E$_{\alpha}$=800 kev, $^{12}$C build up, mixing with Au & Almost BG free condition due to multiplicity \\

Febbraro \cite{Febbraro2020} & 4.2-6.4 &  & University of Notre Dame Nuclear Science Laboratory & E$_{\alpha}$= 1.05, 1.34, 1.59 MeV and $^{27}$Al(p,$\gamma$)$^{28}$Si E$_{p}$=992 keV & EJ315 and EJ301D scintillators & $^{51}$V(p,n)$^{51}$Cr, $^{19}$F($\alpha$,$n$)$^{22}$Na & between 0° and 90° in five-point and twelve-point (near resonances) angular steps & $^{13}$C ACF foils 99\% evap. onto Ta (0.2 mm) & 12-20$\mu$g/cm$^{2}$ & Yield at E$_{\alpha}$= 1.05, 1.34 MeV and E$_{p}$=1.75 MeV & \\

Ciani \cite{Ciani2021} & 0.305-0.4 & 300 $\mu$A & LUNA400, INFN-LNGS & & 18 $^{3}$He based counters embedded in PE with BPE shielding, PSD applied & $^{51}$V(p,n)$^{51}$Cr, AmBe combined with GEANT4 & 4$\pi$ & Enriched $^{13}$C targets using electron gun evaporation onto 0.25 mm Ta backings & 170nm & NRRA $^{13}$C(p,$\gamma$)$^{14}$N at E$_{p}$=1.75 MeV, yield check at E$_{\alpha}$=380 keV & 0.00085(8) c/s and 0.0003(0.3) c/s with PSD $^{c}$ \\

Gao \cite{Gao2022} & 0.31-2.5 & up to 2.5 meA & CJPL and 3 MV Tandetron at Sichuan University & $^{12}$C(p,$\gamma$)$^{13}$N, $^{27}$Al(p,$\gamma$)$^{28}$Si, $^{11}$B(p,$\gamma$)$^{12}$C, $^{14}$N(p,$\gamma$)$^{15}$O & 24 $^{3}$ He filled proportional counters \cite{Yu-Tian2022} & $^{51}$V(p,n)$^{51}$Cr E$_{p}$=1.7-2.6 MeV, GEANT4 & 4 $\pi$ & 2 mm thick $^{13}$C enriched graphite, 97\% & & Repeated yield measurements & 0.0013(5.5) c/s$^{c}$ \\

Brandenburg \cite{Brandenburg2023} & 2.9-8.0 & & Edwards Accelerator
Laboratory at Ohio University & & $^{3}$He and BF$_{3}$ neutron-sensitive proportional counters \cite{Brandenburg2022} & $^{252}$Cf, $^{51}$V(p,n)$^{51}$Cr, $^{13}$C($\alpha$,$n$)$^{16}$O combined with MCNP
& 4$\pi$ & $^{13}$C ACF foils 99\% evap. onto Cu  & (1.12 $\pm$ 0.05) × 1e$^{18}$ atoms/cm$^{2}$ & $\alpha$-elastic scattering, $\alpha$ energy-loss measurements, and scan of the 1.05 MeV $^{13}$C($\alpha$,$n$) resonance & \\

deBoer \cite{deBoer2024} & 0.8-6.5 & 10e$\mu$A & University of Notre Dame Nuclear Science Laboratory & resonances in $^{13}$C($\alpha$,$n$)$^{16}$O & ODeSA, nine deuterated scintillators \cite{Febbraro2019} & $^{9}$Be(d,n)$^{10}$B combined with MCNP simulation & between 0° and 157.5° & similar than Ref. \cite{Ciani2021} & 10.3(6) and $\sim$5 $\mu$g/cm$^{2}$ & NRRA $^{13}$C($\alpha$,$n$)$^{16}$O at E$_{p}$=1.05 MeV & \\

\hline
\end{tabular}
\label{tab:exp_data_summary}
}
\end{center}

\footnotesize $^{a}$ It is referred to Soiné Diplomaarbeit 1991 Stuttgart. \\
\footnotesize $^{b}$ It was stated that \cite{Kellog1989} need to be enhanced with a factor of 1.17.\\
\footnotesize $^{c}$ Beam induced background measurement agrees with environmental background.  \\
\end{sidewaystable*}

\section{Summary}\label{summary} \label{sec:summary}

We have provided an overview of direct measurements of the $^{13}$C($\alpha$,$n$)$^{16}$O reaction cross section and summarised the conclusions of the LUNA experiment, which measured this cross section covering partially for the first time its s-process Gamow window located at $E_{\rm{c.m.}}$=0.14$-$0.25 MeV. 

In spite of the low-energy measurement, extrapolation using direct cross-section data over a wide energy range is still required. Recent and past measurements at higher energy deviate from each other by almost 30\%. Based on our and earlier evaluation, the systematic uncertainty assigned to the neutron detection efficiency and target characterisation can be the main source of the observed discordance.
To further constrain the $S(E)$-factor of the $^{13}$C($\alpha$,$n$)$^{16}$O reaction below $\rm E_{\alpha}$=0.3 MeV, special experimental efforts are needed to reduce systematic and statistical uncertainties over a wide $E_{\alpha}$ range. 

As the systematic uncertainty of neutron detection efficiency is still the main contributor to the uncertainty budget, extension of monoenergetic neutron sources produced by nuclear reactions towards $E_{n}\geq$2 MeV is required, e.g., via the study of the branching of different neutron groups in neutron emitting nuclear reactions ($^{51}$V(p,n)$^{51}$Cr, $^{57}$Fe(p,n)$^{57}$Co, etc.). Another task is to constrain the uncertainties of the angular distribution of the $^{13}$C($\alpha$,$n$)$^{16}$O reaction below $E_{\alpha}$=0.8 MeV. This could be done via dedicated measurements with, e.g., a long counter setup and/or plastic scintillators using a pulsed beam and Time of Flight technique, which could provide data with the required precision and improve the total cross-section calculations.

Moreover, revision of angular distribution and nuclear properties (resonance strength, energy and width) of sharp resonances in $^{13}$C($\alpha$,$n$)$^{16}$O around e.g., $E_{\alpha}$=1.05 and 1.59 MeV can be another way to better constrain the uncertainty of neutron detection efficiency over a wide $E_{\alpha}$ region. 
We have found that in many cases, comparison of simulated and experimental neutron detection efficiency parameters shows discordance. This problem can also be better studied with well known, monoenergetic neutron sources and/or via the study of sharp resonances of $^{13}$C($\alpha$,$n$)$^{16}$O. Proper monitoring of target properties and their modification under beam-bombardment also deserve special attention in future experiments. 

Phenomenological methods for the extrapolation of experimental data, e.g., the R-matrix approach, rely on nuclear physics inputs, such as the parameters of near threshold resonances. Thus, future indirect measurements of, e.g., spectroscopic factors including comprehensive evaluations with old data are needed. As concluded by the most recent work of Ref. \cite{deBoer2024}, multi-channel R-matrix analysis combined with experimental input data and their uncertainty budget is crucial to obtain a more robust low-energy $S$-factor extrapolation. 

\section{Acknowledgement}

Financial support by INFN, the Italian Ministry of Education, University and Research (MIUR) through the "Dipartimenti di eccellenza" project "Science of the Universe", the European Union (ERC Consolidator Grant project STARKEY no. 615604, ERC-StG SHADES no. 852016, and ChETEC-INFRA no. 101008324), Deutsche Forschungsgemeinschaft (DFG, BE 4100-4/1), the Helmholtz Association (ERC-RA- 0016), the Hungarian National Research, Development and Innovation Office (NKFIH K134197), the European Collaboration for Science and Technology (COST Action ChETEC, CA16117) are gratefully acknowledged. UK group acknowledge STFC (grant number ST/V001051/1). For the purpose of open access, the author has applied a Creative Commons Attribution (CC BY) licence to any Author Accepted Manuscript version arising from this submission. M.L. was supported by the the Lend\"ulet Program LP2023-10 of the Hungarian Academy of Sciences. and
NKFIH excellence grant TKP2021-NKTA-64.

\bibliographystyle{model1-num-names}
\biboptions{sort&compress}
\bibliography{13C_review_ref}

\end{document}